\documentclass[sigconf]{acmart}
\AtBeginDocument{%
  }

    \setcopyright{none} % to remove the copyright notice
\usepackage{enumitem}
\usepackage{amsmath,amsfonts}
\usepackage{algorithmic}
\usepackage{graphicx}
\usepackage{textcomp}
\usepackage{xcolor}
\usepackage{listings}
\usepackage{tabularx} % Add this in the preamble
\usepackage[table]{xcolor}
\usepackage{pifont} 
\usepackage{subcaption}
\usepackage{balance}
\usepackage{wrapfig}
\usepackage{booktabs}
\usepackage[utf8]{inputenc}
\usepackage[normalem]{ulem}

\definecolor{codegreen}{rgb}{0,0.6,0}
\definecolor{codegray}{rgb}{0.5,0.5,0.5}
\definecolor{codepurple}{rgb}{0.58,0,0.82}
\definecolor{backcolour}{rgb}{0.95,0.95,0.92}
\definecolor{bblue}{HTML}{4F81BD}
\definecolor{rred}{HTML}{E11916}
\definecolor{ggreen}{HTML}{3FD72D}
\definecolor{ggreen1}{HTML}{9DEC9D}
\definecolor{ppurple}{HTML}{9F4C7C}
\definecolor{yyellow}{HTML}{FFC000}
\definecolor{yyellow1}{HTML}{FEE599}

\definecolor{debug}{HTML}{FFBABA}
\definecolor{info}{HTML}{FF5252}
\definecolor{warning}{HTML}{FF0000}
\definecolor{severe}{HTML}{A70000}

\definecolor{last-year}{HTML}{1E476C}
\definecolor{this-year}{HTML}{9FC5E8}

\lstdefinestyle{mystyle}{
  backgroundcolor=\color{backcolour},   commentstyle=\color{codegreen},
  keywordstyle=\color{magenta},
  numberstyle=\tiny\color{codegray},
  stringstyle=\color{codepurple},
  basicstyle=\ttfamily\footnotesize,
  breakatwhitespace=false,         
  breaklines=true,                 
  captionpos=b,                    
  keepspaces=true,                 
  numbers=left,                    
  numbersep=5pt,                  
  showspaces=false,                
  showstringspaces=false,
  showtabs=false,                  
  tabsize=1,   
  escapeinside={(*@}{@*)}
}

\usepackage{balance}

\usepackage{pgfplots}
\usepackage{tcolorbox}
\usepackage{pgfplotstable}
\usepackage{CJKutf8}  % Enables UTF-8 CJK support

\begin{document}

\title{Evaluating LLM-Based Test Generation Under Software Evolution}
% Do Small Code Changes Matter? A Study of LLM based test generation on Near-Identical Programs
% Evaluating LLM-Based Test Generation Under Software Evolution
%Are LLMs Simply Memorizers When It Comes to Test Generation?

\author{Sabaat Haroon}
\affiliation{%
  \institution{Virginia Tech}
  \country{USA}
}

\author{Mohammad Taha Khan}
\affiliation{%
  \institution{Carnegie Mellon University}
  \country{USA}
}

\author{Muhammad Ali Gulzar}
\affiliation{%
  \institution{Virginia Tech}
  \country{USA}
}
\begin{abstract}

Large Language Models (LLMs) are increasingly used for automated unit test generation. However, it remains unclear whether these tests reflect genuine reasoning about program behavior or simply reproduce superficial patterns learned during training. If the latter dominates, LLM-generated tests may exhibit important weaknesses, including reduced coverage, missed regressions, and undetected faults. Understanding how LLMs generate tests for a program and how those tests respond to code evolution is therefore essential. In this work, we present a large-scale empirical study of LLM-based test generation under program evolution. Using an automated mutation-driven framework, we analyze how generated tests react to semantic-altering changes (SAC) and semantic-preserving changes (SPC). Our evaluation spans eight LLMs and 22,374 program variants derived from widely used benchmarks.

LLMs achieve strong baseline results, reaching 79\% line coverage and 76\% branch coverage with fully passing test suites on the original programs. However, performance degrades as programs evolve. Under SACs, the pass rate of newly generated tests drops to 66\%, and branch coverage declines to 60\%. More than 99\% of failing SAC tests pass on the original program while executing the modified region, indicating residual alignment with the original program behavior rather than adaptation to the updated semantics. Performance also drops under SPCs, despite unchanged functionality: test pass rates fall to 79\% and branch coverage to 69\%. SPC edits typically introduce larger syntactic changes than SAC edits while preserving semantics, yet they trigger greater instability in generated test suites. On average, models produce 1.2× more new tests while discarding many baseline tests, suggesting sensitivity to lexical changes rather than true semantic impact. Overall, our findings show that current LLM-based test generation relies heavily on surface-level cues and struggles to maintain regression awareness as programs evolve.

\end{abstract}

\maketitle

\section{Introduction}

Large Language Models (LLMs) are increasingly integrated into software engineering workflows, with automated test generation emerging as a prominent use case \cite{Fan2023LargeLM}. Early studies and benchmarks suggest that LLMs can generate syntactically valid and often semantically correct test cases for well-known public programming tasks \cite{2026arXiv260212256C, Fuzz4All,10962454,10.1145/3691620.3695513}. However, producing complete and reliable automated test suites requires deep reasoning about control flow, execution paths, and the specific functional state of the provided implementation. Real-world software is inherently dynamic, code is frequently reused, refactored, and slightly tweaked to serve different functional purposes. When developers supply this modified code to an LLM to generate a high-coverage test suite, they expect the resulting tests to accurately reflect the current code.

\noindent\textbf{Problem.} Current studies on LLM-based test generation largely overlook how code evolution impacts model robustness and behavior~\cite{neurosymbolic-oracles-ase2025, 10.1145/3691620.3695513,schafer2023testpilot}. This gap leaves a critical question: do LLMs genuinely comprehend the semantics of the provided code, or are they performing shallow pattern replication of programs seen in pre-training?

Consider a scenario in which a developer takes an existing open-source function and makes a minor semantic-altering modification to adapt it to a new use case. Even when explicitly prompted to generate high-coverage tests for this new code, an ideal test generator (e.g., random test generators~\cite{pacheco2007randoop}, Fuzzers~\cite{manes2019artscienceengineeringfuzzing,Fuzz4All} or Symbolic execution~\cite{cadar2008klee}) should adapt to the provided logic. However, if the LLM relies heavily on memorized structures, it may implicitly assume the modified code is a "buggy" version of the original program. It may also completely overlook the semantic implication of the code change. Consequently, it might generate test cases that fail on the provided code but paradoxically pass on the original, unmodified version seen during training. Conversely, if a developer applies a semantic-preserving mutation, such as renaming variables or refactoring logic without changing the output, the generated test coverage and passing rates should remain stable. If an LLM cannot maintain resilience under these harmless lexical changes, its utility in real-world test generation scenarios is compromised.

\noindent\textbf{Current State of the Art.} Traditional test generation systems~\cite{EvoSuite,lukasczyk2020pynguin,kovalenko2023unittestbot,rho2024tamingbeastfullyautomated}  and fuzzers~\cite{Fuzz4All,manes2019artscienceengineeringfuzzing}, are designed to maximize code coverage, while modern LLM code generation benchmarks (e.g., HumanEval, LiveCodeBench) focus almost exclusively on functional correctness in code generation tasks \cite{10.1145/3597503.3639219}. Neither approach characterizes how an LLM's test generation behavior shifts as software evolves. Although mutation testing is a standard technique for assessing test suite adequacy, it has not been systematically adapted to evaluate LLM-based test generation under program changes. %Without this systematic evaluation, we lack evidence that LLMs can reliably manage test lifecycles. 
It is currently unknown whether LLMs can successfully adapt tests to functional changes while remaining resilient against changes that do not affect behavior.

\noindent\textbf{Contributions.} A challenge in evaluating LLM-based test generation is distinguishing genuine semantic reasoning from alignment with patterns observed during training. High coverage on static benchmarks alone cannot resolve this question. A model may appear effective simply because the program under test resembles implementations encountered during training, allowing it to reproduce familiar test structures without reasoning about the specific code instance. To address this limitation, we evaluate LLM behavior under controlled program changes.

We introduce a mutation-driven evaluation framework that applies two classes of code changes. \emph{Semantic-altering changes} (SACs) produce program variants with modified behavior, requiring tests to adapt to new semantics. \emph{Semantic-preserving changes} (SPCs) modify the code surface while preserving functionality. Because the program behavior remains unchanged, degradation under SPCs reveals sensitivity to structural or lexical variation rather than semantic reasoning. By analyzing LLM behavior under both change types, we isolate whether performance differences arise from semantic misunderstanding or reliance on superficial code patterns. This dual perspective allows us to characterize three properties of LLM-based test generation: sensitivity to semantic change, resilience to non-functional structural change, and stability of generated test suites across evolving programs.

We design an automated end-to-end evaluation framework that generates baseline tests, measures coverage, injects SACs and SPCs, and evaluates test quality across program variants. Beyond coverage metrics, we analyze failing tests to determine whether their assertions align with the mutated program behavior or with the original program semantics. This analysis enables us to directly assess whether LLM-generated tests reflect true reasoning about program behavior or shallow pattern replication---insights that static public benchmarks cannot provide.

\noindent\textbf{Experimental Results.} We evaluated 8 state-of-the-art LLMs using 22,374 program variants sourced from the Project CodeNet dataset~\cite{puri2021codenet}, specifically focusing on Java and Python implementations. Our baseline evaluation reveals that LLMs perform well on the original, unmodified programs, achieving an average of 79.2\% line coverage and 76.1\% branch coverage with fully passing test suites, containing 13.1 tests per program, on average. However, this performance degrades sharply under code changes. When subjected to SACs, the pass rate of newly generated tests plummets to 66.5\%, and branch coverage falls to 60.6\%. Crucially, further analysis of failing test cases finds that, among 119,163 tests generated under SACs, over 99\% of the 23,977 tests that failed on the mutated code pass on the original program when executing the modified region. These results support our hypothesis that \emph{LLMs lack the precision to reason about semantic code changes and instead rely heavily on patterns observed during training}. As a result, they often fail to capture behavioral changes introduced by evolving programs and rarely generate tests that reflect the updated semantics. 
%\gulzar{We do not know if LLM is training the change as a bug or it didn't pick the change, so let's not say that. Say that it is fixated on the original program. }\sabaat{Addressed}

Furthermore, when subjected to SPCs, the test pass rate decreases from 100\% to 79\%, and branch coverage drops to 69\%. Despite the same functionality, LLMs interpret SPC-modified code as substantially different programs, triggering 1.2$\times$ more new test generations than under SACs. We attribute this behavior to the slightly larger syntactic changes introduced by SPCs, which typically span 1-2 lines of code, whereas SACs are restricted to single-line changes. As a result, changes with higher syntactic impact but no semantic effect lead to larger differences in generated test suites, while changes with real semantic impact but minimal syntactic change produce smaller shifts. However, a model capable of meaningful code reasoning should remain robust to such SPCs. Instead, our findings show that \emph{larger edit distances---even when semantically irrelevant---disrupt LLMs' pattern matching and trigger exploratory regeneration of tests, leading to unstable test suites}. Overall, these findings indicate that LLM-based test generation relies heavily on surface-level patterns and memorized structures, revealing a limited ability to comprehend and adapt to evolving code.

\noindent \textbf{Data Availability.} The artifacts and datasets used in this study are publicly available at \url{https://doi.org/10.5281/zenodo.18898624}.

\section{Motivating Example}
\label{sec:motivating_example}

%To illustrate the unpredictable nature of LLM-based test generation under code, 
We present a motivating example based on an instance from the CodeNet Python dataset (ID: \texttt{p02701\_s025427336}). 
The program reads an integer \(n\) followed by \(n\) strings, groups them by length (from 1 to 10 characters), and prints the total number of distinct strings across all lengths. 
We first prompt \textit{GPT-5.2} to generate a test suite for the original, unmodified program. The model successfully generates a suite of 7 test cases, achieving 100\% line coverage and 81.0\% branch coverage, with a 100\% test pass rate. %\gulzar{count of tests? and which LLM?}\sabaat{Addressed}
%\gulzar{creating 100\% passing tests does not mean good understanding, it means that LLMs have understood the execution to generate input that leads to high coverage. We need to insist on coverage as a test objective and should clarify that upfront}\sabaat{Addressed}
This establishes a strong baseline, demonstrating the problem's tractability and that LLMs can meet high code-coverage objectives by generating executable, passing test cases for the string-bucketing logic.

Next, we simulate a functional extension by injecting a code change that alters semantics. 
Specifically, we modify the input-reading logic to process only the first $n-1$ entries, effectively treating the final input as a terminator or metadata rather than a data point. This change establishes a new valid functional requirement: the program intentionally ignores the last string in the sequence. Note that we do not aim for functional alignment, as LLMs are not provided with the code's functional requirements. Instead, LLMs are asked to achieve code coverage as the primary objective, as in the automated test generation literature~\cite{neurosymbolic-oracles-ase2025,schafer2023testpilot}. 

We then provide this modified version to a separate, independent instance of the LLM, instructing it with generating a high-coverage test suite from scratch. By using a fresh session, we ensure that the model has no memory of the previous baseline code, forcing it to rely entirely on the provided code to reason upon.

\begin{figure}[t]
    \centering
    \begin{lstlisting}[language=Python, numbers=left, numberstyle=\tiny, stepnumber=1, numbersep=3pt,
basicstyle=\ttfamily\scriptsize, keywordstyle=\color{blue}, commentstyle=\color{codegreen}, 
stringstyle=\color{red}, breaklines=true, frame=none, xleftmargin=1em]
n = int(input())
goods1, goods2 = [], [] (*@\textcolor{gray}{\# ... up to goods10}@*)
count = 0

(*@\textcolor{purple}{\# SAC: Shifted to exclusive range logic}@*)
for _ in range((*@\textcolor{red}{n - 1}@*)): (*@\textcolor{gray}{\# originally range(n)}@*)
  check = input()
  finder = len(check)
  if finder == 1:
    if check not in goods1:
      goods1.append(check)
      count += 1
  (*@\textcolor{purple}{\# SPC: Redundant Else injected here}@*)
  (*@\textcolor{blue}{else:}@*)
      (*@\textcolor{blue}{pass}@*)
      
  if finder == 2:
(*@\textcolor{gray}{\# <ADDITIONAL CODE ON REMAINING LENGTH CHECKS>}@*)
print(count)
\end{lstlisting}
% \vspace{-3ex}
    \caption{Fragment of CodeNet program showing injected SAC and SPC}
    \label{fig:motivating-example-codenet}
    % \vspace{-3ex}
\end{figure}

When evaluating the tests generated for this mutant, we observe that
line coverage drops moderately to 68\%, and branch coverage falls to 50\%. However, the test pass rate drops to 82\%, meaning several tests fail. 
Upon closer inspection, the specific failing test case shown in Figure~\ref{fig:motivating-test-case} stands out as particularly revealing: \texttt{test\_all\_lengths\_unique}.
This test generates inputs that feed exactly 10 distinct strings through the bucketing logic. However, despite receiving no external specification in the prompt, the model explicitly asserts that the program will output \texttt{"10"}. This reveals that rather than deriving the expected behavior from the provided mutated code, the LLM forcefully matches the assertion to the original algorithmic specification it observed during pre-training.

%\gulzar{This might be problematic, because during test generation, we do not provide the expected spec to LLM, so how does LLM know to check that? We need to argue this in terms of coverage and say that even when we do not provide spec, it tries to match the code's spec with what it had seen previously.}\sabaat{Addressed}
\begin{figure*}[t]
  \centering
  \includegraphics[width=\textwidth,keepaspectratio]{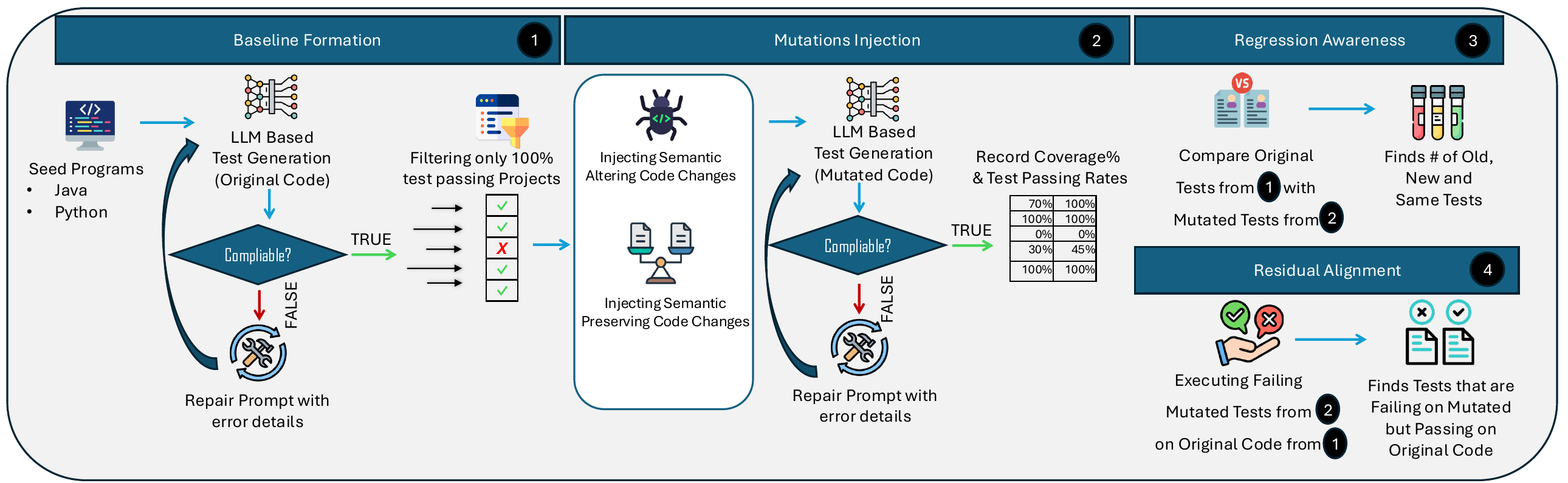}
  % \vspace{-6ex}
  \caption{Methodology Overview}
  \label{fig:overview}
  \vspace{-2ex}
\end{figure*}
\begin{figure}[t]
    \centering
    \begin{lstlisting}[language=Python, numbers=left, numberstyle=\tiny, stepnumber=1, numbersep=3pt,
basicstyle=\ttfamily\scriptsize, keywordstyle=\color{blue}, commentstyle=\color{codegreen}, 
stringstyle=\color{red}, breaklines=true, frame=none, xleftmargin=1em]
def test_all_lengths_unique(self, capsys):
    """Test that strings of all lengths 1-10 are correctly counted."""
    inputs = ["10", "a", "bb", "ccc", ..., "jjjjjjjjjj"]
    mod = self._import_module_with_input(inputs)
    out = capsys.readouterr().out.strip()
    (*@\textcolor{purple}{\# Fails under SAC: out is "9" instead of "10"}@*)
    assert out == "10", "All 10 unique strings should be counted"
    assert mod.goods10 == ["jjjjjjjjjj"]
\end{lstlisting}
% \vspace{-3ex}
    \caption{The intriguing \texttt{test\_all\_lengths\_unique} test case generated by the LLM}
    \label{fig:motivating-test-case}
    % \vspace{-3ex}
\end{figure}

On the original program, this test passes. When subjected to SAC the loop terminates after 9 iterations, the 10th string (\texttt{"jjjjjjjjjj"}) is never read, and the output is \texttt{"9"}, causing the assertion to fail. 
%While such a test would be ideal for a version intended to process $n$ strings, it is logically incorrect for this specific program. 
The LLM’s failure to adjust the expected output to \texttt{"9"} demonstrates a lack of semantic grounding. Rather than deriving assertions from the provided code, the model displays \emph{residual alignment}: it hallucinates the requirements of the standard algorithm from its training data, ignoring the explicit functional logic of the current instance. 
%\gulzar{This is a good reasoning}
%

%This proves that under semantic-altering code changes, LLM test generation becomes misaligned with the actual code semantics, yielding moderate coverage drops and unexpected test failures.

%%
To further evaluate robustness, we return to the original baseline code and apply a semantic-preserving code change (SPC). 
We insert a purely redundant \texttt{else: pass} block immediately following the first length check (see Figure~\ref{fig:motivating-example-codenet}). 
Because this change preserves the exact same control flow and program behavior, a robust LLM should generate a test suite comparable to the baseline. 
Instead, the SPC causes line coverage to drop to 79\% and branch coverage to 64\%. More concerningly, the test pass rate drops to 80\%. 

This degradation reveals an over-reliance on syntactic visibility at the expense of semantic reasoning. Because SPCs often introduce highly visible structural modifications such as inserting an entirely new code block, the LLM easily detects the surface-level disruption but struggles to recognize its semantic impact, prompting it to unnecessarily discard valid tests and generate new, inferior suites. Conversely, SACs typically possess a minimal syntactic footprint (e.g., modifying a single operator) but carry larger semantic implications. Because these behavioral shifts are visually subtle, LLMs frequently overlook them. %Together, this contrast demonstrates that LLM-based test generators act as superficial pattern matchers rather than robust semantic analyzers.

%\gulzar{Please fill this in with the logic about how the SPCs are more visible and thus detectable, but LLMs are incapable of accurately assessing their semantic impact. However, for SAC, they are small yet have a high semantic impact, but LLMs do not pick that. 

%These contrasting behaviors, where the LLM is simultaneously insensitive to meaningful logical updates, yet hypersensitive to non-functional structural refactoring are two symptoms of the exact same underlying limitation: a reliance on shallow pattern matching rather than deep semantic reasoning. Under Semantic-Altering Code Changes, strong pre-training priors overpower the explicit prompt, blinding the model to deliberate functional shifts as it defaults to residual alignment. Conversely, under Semantic-Preserving Code Changes, harmless lexical variations disrupt the exact token sequences the model relies upon to trigger its generation capabilities, breaking its fragile pattern recognition and causing test quality to collapse.  This dual vulnerability, where LLMs fail to track intentional change in functionality while drastically overreacting to benign structural refactoring, serves as the core motivation for our systematic evaluation framework. \gulzar{The two reasoning sorts of contradict. In one case, we say that an LLM cannot detect small changes, and in the other, we say that it is highly sensitive to them. }\sabaat{Addressed}
\section{Research Questions}
\begin{table*}[ht]
    \centering
    \resizebox{\textwidth}{!}{%
    \begin{tabular}{l l p{8cm} l}
    \hline
    \textbf{Modification} & \textbf{Type} & \textbf{Description} & \textbf{Example} \\
    \hline
    \multicolumn{4}{c}{\textit{Semantic-Altering Code Changes(SACs)}} \\
    \hline
    Boundary Shift & Semantic-Altering & Modifies the inclusive or exclusive range of a loop or array index. & \texttt{range(n)} $\rightarrow$ \texttt{range(n-1)} \\
    Changed Boolean Logic & Semantic-Altering & Switches boolean operators to alter decision logic. & \texttt{a \&\& b} $\rightarrow$ \texttt{a || b} \\
    Changed Arithmetic & Semantic-Altering & Switches arithmetic operators to alter calculation values. & \texttt{a + b} $\rightarrow$ \texttt{a - b} \\
    Argument Swap & Semantic-Altering & Reorders arguments of the same type in a function call, shifting the functional role of the inputs within the program logic. & \texttt{f(sum, price)} $\rightarrow$ \texttt{f(price, sum)} \\
    Variable Role Rebinding & Semantic-Altering & Reassigns a variable's logical role. & \texttt{for x in data: total += x} $\rightarrow$ \texttt{for total in data: total += total} \\
    \hline
    \multicolumn{4}{c}{\textit{Semantic-Preserving Code Changes(SPCs)}} \\
    \hline
    Void Loop Injection & Structural & Inserts dummy loops that execute but perform no operation. & \texttt{for i in range(1): pass} \\
    Void Condition & Structural & Inserts dummy conditions that always execute but do nothing. & \texttt{if (true) \{ ... \}} \\
    Redundant Else & Structural & Adds empty \texttt{else} blocks to statements lacking them. & \texttt{if x: y else: pass} \\
    Equivalent Comparison & Structural & Rewrites comparison operators to logically equivalent forms. & \texttt{x < 10} $\rightarrow$ \texttt{!(x >= 10)} \\
    Unused Parameter & Structural & Adds an unused parameter to method signature and updates calls. & \texttt{def f(x):} $\rightarrow$ \texttt{def f(x, null):} \\
    Misleading Variables & Identifier & Renames variables to generic or misleading names. & \texttt{count} $\rightarrow$ \texttt{sum} \\
    Misleading Comments & Annotative & Adds comments that misdescribe code behavior. & \texttt{// Returns sum} (above diff code) \\
    Misleading Mandarin & Annotative & Adds misleading comments translated into Mandarin Chinese. & \texttt{// \begin{CJK*}{UTF8}{gbsn}返回总和\end{CJK*}} \\
    Remove Comments & Annotative & Removes all single-line and multi-line comments. & \texttt{/* comment */} $\rightarrow$ \textit{(empty)} \\
    \hline
    \end{tabular}%
    }
    \caption{Types of Semantic Altering and Semantic Preserving Code Changes Applied to Seed Programs}
    \label{tab:mutations}
\end{table*}
\label{RQs}
To guide this study of LLM-based testing under the program
changes, we investigate the following research questions:
\begin{itemize}
\item \textbf{RQ1:} To what extent can LLMs generate structurally valid and high-coverage test suites for software benchmarks in their initial state?
\item \textbf{RQ2:} How sensitive are LLMs to semantic-altering code changes, where meaningful changes should cause tests to adapt?
\item \textbf{RQ3:} How resilient are LLMs to semantic-preserving code changes, where program behavior is functionally identical?
\item \textbf{RQ4:} What explains the failures of LLM-generated tests under Semantic Altering Code Changes?
\item \textbf{RQ5:} To what extent do LLM-generated test suites demonstrate regression awareness?
\end{itemize}

\section{Methodology}
%The goal of this work is to rigorously evaluate how software changes  affect the reliability of LLM-based test generation.  %\gulzar{We need to clarify why we refer to mutations as "changes." The mutations we are discussing are not representative of actual changes found in the real world. In software development, changes typically refer to commits or pull requests made by individuals. While these changes are indeed better than mutations, they may not be feasible for our purposes because (1) LLMs have already been exposed to those changes, as well as the corresponding regression and other tests associated with them. and (3) creating them manually is not scalable.}\sabaat{Addressed} 
%To achieve this, 

%This approach guarantees that the model is evaluated on strictly unseen logic. Using this paradigm, we investigate whether models can autonomously adapt test assertions to reflect new functional specifications and maintain consistent test quality when faced with structural modifications that preserve original program behavior.

%Because language models occasionally produce malformed outputs on the first attempt, our framework incorporates a single-iteration repair mechanism. If the initially generated test suite contains syntax errors or fails during execution, we extract the resulting error trace and provide it back to the LLM, granting it one subsequent chance to repair the suite. 

We design an automated pipeline that benchmarks the adaptability of LLM-generated test suites. As shown in Figure \ref{fig:overview}, our framework subjects the model to both SACs and SPCs to quantify its ability to remain aligned with the current code state.  In the first phase, we establish a baseline to assess the model's initial generation capabilities. We prompt the LLM to generate test suites for a set of  seed programs and strictly retain only those programs for which all generated tests compile and pass perfectly. 
%This baseline phase is crucial to ensure that the LLM is fully capable of generating valid test cases and that the programming task is not ambiguous. By filtering out programs where the LLM immediately fails to produce correct syntax or executable tests, we isolate the root cause of subsequent failures. 
It ensures that any degradation observed in later stages is directly attributable to the injected code changes, rather than a general inability to comprehend the original codebase. The second phase evaluates the model's adaptability to code changes.
We apply SACs, detailed in Table~\ref{tab:mutations}, to represent functional changes, such as changing a loop boundary or an arithmetic operator.  Second, we apply SPCs, also outlined in Table~\ref{tab:mutations}, to represent non-functional code refactoring, such as inserting a redundant condition or renaming a variable.  While real-world commits and pull-requests reflect natural developer behavior, utilizing them introduces a critical issue of data contamination i.e., LLMs have already been exposed to these open-source commits, along with their corresponding regression tests, during pre-training. Furthermore, manually crafting novel, realistic pull requests at scale is infeasible. Therefore, systematically applied mutations serve as a scalable and robust proxy for code changes. 

To facilitate a uniform evaluation, we employ a standardized two-shot prompting strategy, where the second shot performs a repair cycle for the LLM in cases where the generated tests fail to compile due to errors, and this strategy is applied consistently across all phases of the pipeline. The prompt instructs the model to act as an expert developer and generate a high-coverage, successfully passing test suite based entirely on the provided code snippet. Instead of a custom prompt design, we employ prompting strategies that are commonly used in prior LLM-based unit test generation studies, where the model is instructed to generate tests directly from the provided implementation\cite{yuan2023chatunitest,schafer2023testpilot}. If the tests remain erroneous after 2-shot attempt, the test generation process for that specific program version is marked as a failure, and the program is excluded.

% First, The expected behavior is that a semantically grounded LLM will detect this logical shift and generate updated test assertions to cover the new functionality. Because the program's functional behavior remains identical, the expected behavior is that the LLM will maintain its baseline test coverage and pass rates without shedding valid tests. This dual-phase methodology allows us to rigorously evaluate whether test generation failures stem from an inability to reason about changing program semantics or a fragile overreliance on superficial code structure.
% \gulzar{In evaluation, we should add number on programs on which LLM was not able to generated tests correctly leading to syntactic/runtime errors}\sabaat{will do}

% This design allows us to isolate failures caused by inadequate reasoning about program semantics from failures caused by task ambiguity or underspecification.

\subsection{Seed Programs Procurement}
We procure our seed programs from the \textbf{Project CodeNet} dataset~\cite{puri2021codenet}, a large scale, widely recognized benchmark. We focus our empirical study on \textbf{Java} and \textbf{Python}, two widely used programming languages heavily represented in both real-world software systems and LLM pretraining corpora. By initializing our pipeline with \textbf{5,723} candidate programs (\textbf{3,231 Java} and \textbf{2,492 Python}) sourced from CodeNet, we evaluate models at a scale that significantly exceeds standard test generation benchmarks, such as HumanEval (164 problems)\cite{HumanEval} or MBPP (974 problems)\cite{MBPP}. This large scale selection ensures our findings are robust and builds upon a verified dataset successfully utilized in numerous previous empirical software engineering studies~\cite{khajezade2024investigating,nicoletti2024crosslingual}. We follow the following criteria.  %To ensure a valid and reliable evaluation baseline, we enforce a strict quality filter across the procured CodeNet dataset. Every selected program must satisfy critical criteria to isolate the model's test generation capabilities:

\begin{itemize}
    \item \textbf{Dataset Criteria:} We enforce a strict quality filter requiring all selected programs to be natively compilable, executable, and strictly self-contained. This ensures the model receives the complete logical state without relying on external dependencies.
    \item \textbf{Scale:} We initialize our pipeline with \textbf{5,723} candidate programs (\textbf{3,231 Java} and \textbf{2,492 Python}).
    \item \textbf{Stratified Sampling:} We partition the dataset into four systematic strata based on Lines of Code (LOC) percentiles: 0-25\%, 25-50\%, 50-75\%, and 75-100\%. By defining the boundaries using these quartiles, each band contains an identical number of candidate projects rather than relying on arbitrary line count limits. Uniformly sampling passing projects from these equal-frequency bands guarantees a balanced evaluation across all code complexities and prevents our results from skewing toward small or large files.
\end{itemize}

\subsection{Baseline Test Generation and Program Filtering}

\begin{table}[t]
  \centering
  \scriptsize
  \setlength{\tabcolsep}{6pt}
  \renewcommand{\arraystretch}{1.2}
  \setlength{\aboverulesep}{0pt}
  \setlength{\belowrulesep}{0pt}
  \begin{tabular}{l l c c}
    \toprule
    \textbf{Model} & \textbf{Creator} & \textbf{Size} & \textbf{Type} \\
    \midrule
    GPT-OSS \cite{gptoss}                 & OpenAI    & 20 B        & Open-source \\
    Nemotron-3-Nano \cite{nemotron3nano}  & NVIDIA    & 30 B        & Open-source \\
    GPT-5 \cite{gpt5}                     & OpenAI    & Undisclosed & Closed-source \\
    GPT-5.2 \cite{gpt52}                  & OpenAI    & Undisclosed & Closed-source \\
    Claude 4.5 Haiku \cite{claude45haiku} & Anthropic & Undisclosed & Closed-source \\
    Claude 4.6 Sonnet \cite{claude46sonnet}& Anthropic & Undisclosed & Closed-source \\
    Gemini 2.5-Flash \cite{gemini25flash} & Google    & Undisclosed & Closed-source \\
    Gemini 3.1-Pro \cite{gemini31pro}     & Google    & Undisclosed & Closed-source \\
    \bottomrule
  \end{tabular}
  \caption{LLMs Evaluated}
  \label{tab:llms}
  \vspace{-4ex}
\end{table}

For each seed program, we instruct a set of 8 LLMs (listed in Table~\ref{tab:llms}) to generate high-coverage test suites targeting the original code's logic. These models were selected for their prominence in current literature and include both widely used open-source and proprietary alternatives. Open-source models were executed on local servers equipped with NVIDIA L40S GPUs, while closed-source models were accessed via their respective APIs. Once generated, the test suites are executed against the original programs to establish a baseline for correctness and code coverage.

We systematically limit our final baseline set to exactly 100 fully passing projects per LLM. Using the four line-count bands defined during procurement, we continuously generate and evaluate baseline test suites until we successfully identify 25 passing projects from each of the four size bands. This stratified sampling guarantees our evaluation is thoroughly balanced across varying code complexities and is not skewed by a disproportionate number of trivial snippets or excessively large files.  Crucially, we only select 100 programs because each baseline program is independently subjected to 14 distinct code changes (comprising all categories of SACs and SPCs), an initial set of 100 baseline programs expands to nearly 1,400 independent test suites per evaluated LLM. This massive multiplier effect ensures a statistically rigorous, large-scale evaluation of code evolution while remaining computationally feasible across 8 different models.

%The primary objective of our study is to measure the semantic sensitivity of LLM test generation under code changes, rather than merely evaluating a model's baseline ability to write tests. Therefore, 

We enforce a strict filtering condition: a program is only selected for the final 100-project baseline if the LLM successfully generates a test suite with a 100\% pass rate on the original code. Any program for which the model produces syntax errors, compilation failures, or at least one failing test is discarded.  If a model fundamentally fails to comprehend the original program to generate valid tests, evaluating its response to a subsequent code changes yields confounded results.

This rigorous filtering step creates a flawless control group. It guarantees that the LLM demonstrates absolute baseline competence for every retained program. Consequently, we can definitively attribute any subsequent degradation in test pass rates or coverage observed in the later stages directly to the injected mutations, completely isolating the impact of code evolution from inherent task ambiguity or baseline generation deficits.

\subsection{Semantic-Altering Code Changes}
We apply \emph{Semantic-Altering Code Changes} to the filtered seed programs. These changes represent functional divergences from the original code, simulating common scenarios where a program's logic is updated to meet new operational requirements or boundary conditions. In this phase, we provide the modified code to an independent instance of the LLM and instruct it to generate a high-coverage test suite. This generation process ensures the model treats the code as a primary reference, allowing us to evaluate its semantic alignment without the influence of prior versions. This allows us to measure how semantic changes affect the test generation, including changes in test suite size, structural diversity, and code coverage.

%\gulzar{refer to table and give two examples of how SAC works}\sabaat{Addressed}
Table~\ref{tab:mutations} lists the SACs used.
%, SACs systematically alter program behavior by introducing atomic functional shifts. 
Consider two specific examples from our methodology. First, a \emph{Boundary Shift} code change changes a loop condition from \texttt{range(n)} to \texttt{range(n-1)}. The expected behavior is that the LLM will generate tests reflecting one fewer iteration in the expected output. Second, a \emph{Changed Arithmetic} mutation swaps an addition operator (\texttt{a + b}) for a subtraction operator (\texttt{a - b}), obligating the LLM to generate assertions that check for a completely different mathematical result. %We focus on code changes that lead to atomic functional shifts to establish a baseline for the LLM's semantic sensitivity. If a model fails to align its testing logic with these fundamental changes to operators or boundaries, it indicates a significant lack of grounding in the provided source code, making it unsuitable for more complex, multi-line logic updates.

\subsection{Semantic-Preserving Code Changes}

To evaluate the robustness of LLM-based test generation, we apply \emph{Semantic-Preserving Code Changes} (SPCs) to the filtered baseline programs. Real-world software maintenance frequently involves code refactoring, variable renaming, and documentation updates that do not alter the underlying execution logic. A reliable, production-ready testing agent must maintain consistent performance across these functionally identical variations. %We apply SPCs to rigorously measure whether models truly understand the execution semantics or if they merely pattern-match against familiar structures observed during training.

%Following the same protocol as our semantic-altering evaluation, 
We provide the refactored code to an independent instance of the LLM and instruct it to generate high-coverage tests.
%This strict isolation prevents the model from relying on prior conversational context, forcing it to analyze the modified snippet as a standalone implementation. 
Because SPCs strictly preserve program semantics, the expected behavior is that LLM will recognize the unchanged execution logic and generate a test suite that achieves coverage and pass rates comparable to the baseline scenario. Any degradation in coverage or passing\% serves as empirical evidence that the model is over-reliant on superficial syntactic or lexical cues rather than active code comprehension.

Table~\ref{tab:mutations} lists SPCs. Across our mutation pipeline, SACs modify an average of 1.0 line per program, whereas SPCs modify 2.4 lines on average (with some affecting up to 7 lines); we later analyze how this larger syntactic edit size influences test-suite stability (Section~\ref{rq5results}). For SPCs, consider two specific examples. First, an \emph{Equivalent Comparison} mutation rewrites a conditional statement from \texttt{x < 10} to its logical equivalent \texttt{!(x >= 10)}. The expected behavior is that the LLM will recognize the identical boundary condition and generate the exact same test assertions. Second, a \emph{Misleading Variables} mutation might rename a simple loop counter from \texttt{count} to \texttt{sum}. A robust model should derive its understanding from the actual operations applied to the variable, rather than being tricked by the deceptive identifier into generating irrelevant assertions about addition. We simulate SPCs into three distinct groups, with full details provided in Table~\ref{tab:mutations}:

\begin{itemize}
    \item \textbf{Structural Refactoring:} We introduce control-flow transformations that strictly preserve behavior (e.g., injecting a \emph{Void Loop} or \emph{Redundant Else}) to test whether the LLM becomes disoriented by the presence of harmless, non-operational execution paths.
    \item \textbf{Identifier Refactoring:} We systematically rename variables to generic or contextually deceptive identifiers to evaluate whether the model relies heavily on semantic naming conventions rather than programmatic logic.
    \item \textbf{Annotative Refactoring:} We modify, add, or remove comments to manipulate the model's textual context, including injecting cross-lingual \emph{Misleading Mandarin} comments, to assess if logical reasoning can be derailed by textual artifacts that contradict actual code behavior.
\end{itemize}
% \gulzar{Section 4 is entirely written as an explanation of steps. What you need to focus on is why each step is needed, how it helps solve or evaluate an RQ, and what the expected or anticipated behavior. Please give examples.}\sabaat{Addressed}
\subsection{Failure Attribution Analysis}

Consider RQ4 from Section~\ref{RQs}, we extend our evaluation pipeline with a failure attribution analysis applied to tests generated by LLMs on programs with SACs. For each SAC, we collect all LLM-generated tests that fail when executed on the changed program. For each such failing test, we perform two additional checks:

\paragraph{Original-Program Execution.}
The failing test is executed on the original, unmodified version of the program. If the test also fails on the original program, the failure is attributed to poor or malformed test generation rather than misalignment with program's logic.
\begin{figure*}[t]
    \vspace{-1ex}
    \centering
    \includegraphics[width=0.9\textwidth,keepaspectratio]{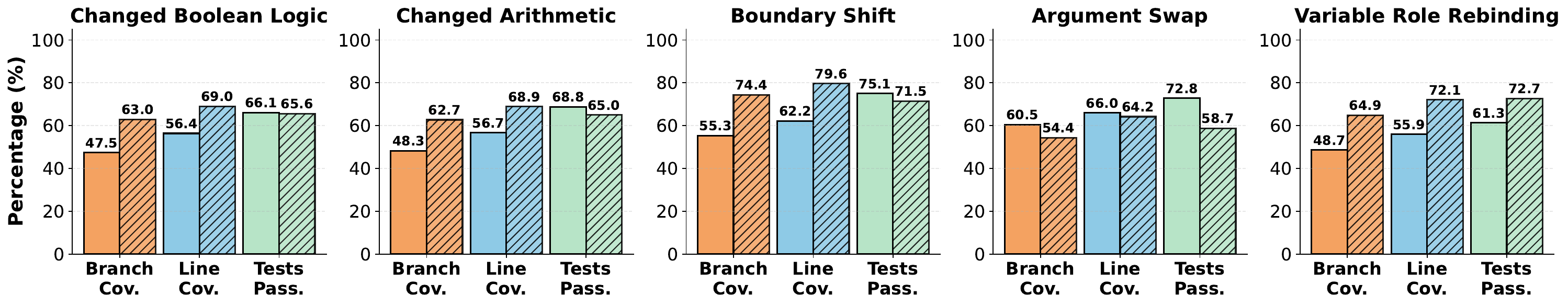}
    \includegraphics[width=0.5\textwidth,keepaspectratio]{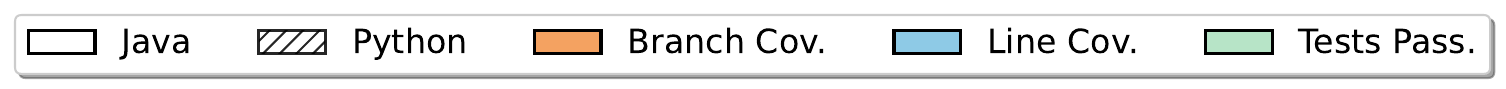}
    \caption{Performance degradation of all LLM-generated test suites across Semantic-Altering Code Changes}
    \label{fig:sam-coverage}
    \vspace{-2ex}
\end{figure*}
\paragraph{Code Change Execution Coverage.}
We verify whether the test executes the changed statement or control-flow region, ensuring that the observed failure is causally related to the code change rather than an unrelated execution path. A failing test is classified as \emph{residually aligned with the original program semantics} if it satisfies both of the following conditions: (1) it passes when executed on the original program, and (2) it executes the mutated code region while failing on the mutated program. This analysis validates our hypothesis that LLMs tend to match code to specifications resembling programs seen during training. Consequently, they often overlook code changes with substantial semantic impact and generate tests that remain aligned with the original program behavior, even when the modified program exhibits diverging semantics.

% \gulzar{What is the difference between 4.7 and 4.3? If not needed, turn this into an implementation detail section, explaining the details of the code. and APIs. }\sabaat{Addressed, I added 4.7 just to emphasize on the point that we are taking only 100\% passing test cases to phase 2 so any degradation in coverage or test passing\% can only be attributed to changes but now I have incorporated this point in 4.3}
% \subsection{Filtering Baseline Programs}
% Our study is designed to measure the semantic sensitivity of LLM-based test generation. Rather than evaluating the model's baseline ability to produce a test suite, we focus on whether its reasoning remains the same in the current source code as it undergoes functional and structural changes. To establish a reliable baseline, we therefore restrict our analysis to programs for which an LLM is able to generate a fully passing test suite for the original, unmodified code. This filtering step ensures that the model demonstrates adequate test-generation capability for the original program, allowing us to attribute any subsequent degradation in test passing percentage or coverage directly to the effects of code evolution. By conditioning on successful baseline performance, our evaluation isolates the impact of semantic-altering and semantic-preserving code changes on LLM behavior.

\section{Experimental Results}

This section presents the empirical results of our large-scale evaluation of LLM-based test generation under code changes. 
%Our analysis focuses on how different categories of code changes impact test generation effectiveness, measured in terms of test pass rate, line coverage, branch coverage, and the number of generated tests. 
Across the entire evaluation pipeline, we ran $22,374$ test generation tasks on the evaluated models, consuming approximately $346$ million tokens (including both input prompts and generated outputs).

\subsection{RQ1: Baseline Performance on Unmodified Programs}
We begin by establishing a control. For each seed program, we prompt the LLMs to generate test suites for the original, unmodified code. To ensure we are evaluating the impact of mutations rather than the inherent model's ability to generate tests for that program, we strictly retain only those programs for which the generated tests achieve a 100\% pass rate. 

\begin{figure}[t]
  \centering
  \includegraphics[width=0.95\columnwidth]{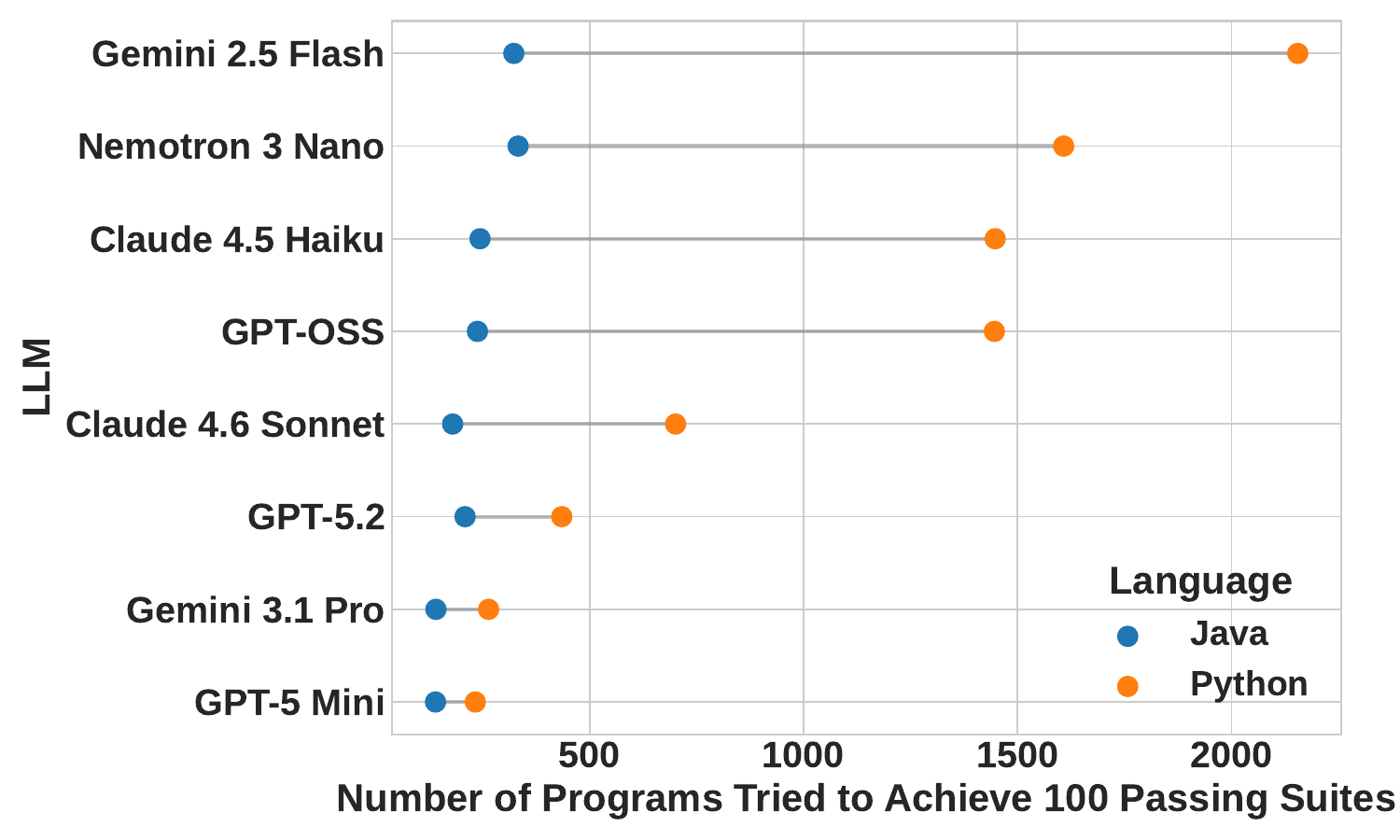}
  \caption{Programs evaluated per model to achieve exactly 100 fully passing baseline test suites. The horizontal gap highlights the increased difficulty of test generation in Python compared to Java.}
  \label{fig:baseline_effort}
\end{figure}

While this filtering step guarantees a control group for later phases, tracking the number of attempts required to secure these 100 passing programs per model reveals significant variations in baseline test generation capabilities. As illustrated in Figure~\ref{fig:baseline_effort}, the efficiency of LLMs varies drastically across both models and programming languages. Highly capable models like GPT-5, Gemini 3.1 Pro, and Claude 4.6 Sonnet required relatively few attempts to hit the target threshold (e.g., Claude 4.6 Sonnet tried 180 programs to secure 100 passing Java suites). Conversely, we observe a severe performance bottleneck in Python test generation across several models. For example, Gemini 2.5 Flash required 2,155 attempts in Python compared to only 323 in Java. This indicates that producing fully self-contained, syntactically correct, and logically sound test suites in Python poses a distinctly higher baseline challenge for current LLMs. 

We attribute this discrepancy primarily to the fundamental differences in the languages' type systems. Java's static and strong typing enforces explicit type declarations for all variables, method arguments, and return types. This rich syntactic structure provides the LLM with definitive deterministic context, dramatically reducing the search space for valid test inputs and expected behaviors. In contrast, Python's dynamic typing requires the model to implicitly infer expected data types and object structures from the surrounding control flow. This reliance on implicit type inference forces the LLM to make frequent assumptions, significantly increasing the probability of generating type-mismatched inputs, invoking incompatible methods, or hallucinating incorrect object structures during test generation. Beyond type systems, this performance gap may also stem from a scarcity of formal Python tests in the models' training corpora. Python is heavily utilized in exploratory environments such as computational notebooks, where developers rarely write structured unit tests~\cite{Nguyen_2025}.
\begin{figure*}[ht]
    \vspace{-2ex}
    \centering
    \includegraphics[width=1\textwidth,keepaspectratio]{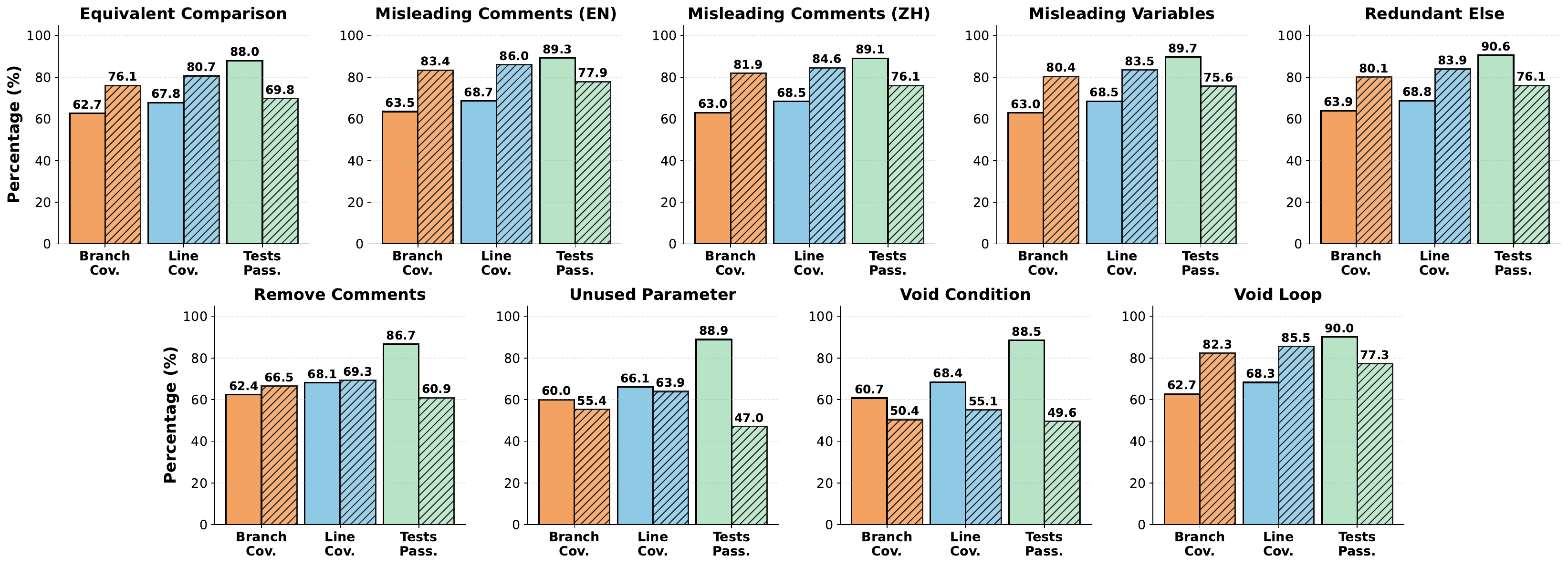}
    \includegraphics[width=0.5\textwidth,keepaspectratio]{figures/results_by_SPM_type_legend.pdf}
    \caption{Performance degradation of all LLM-generated test suites across Semantic-Preserving Code Changes}
    \label{fig:spm-coverage}
    \vspace{-2ex}
\end{figure*}

\subsubsection*{Analyzing Baseline Test Failures}
Analyzing the discarded programs sheds light on where LLMs struggle before any code changes are even introduced. The failures that prevented test suites from achieving a 100\% pass rate generally fall into two categories:
\begin{itemize}
    \item \textbf{Syntactic and Environmental Errors:} A significant portion of failures i.e., 39\% of failures, particularly in Python, resulted from structural hallucinations. Models frequently generated tests that attempted to import nonexistent testing utilities or hallucinated external file dependencies that violated our self-contained dataset criteria. In Java, failures often involved incorrect class instantiation or mismatched accessibility modifiers (e.g., attempting to test private helper methods without reflection).
    \item \textbf{Logical Assertion Failures:} In 61\% of discarded cases, the tests compiled successfully but failed during execution because the LLM fundamentally misunderstood the program's edge cases. These logical mismatches prove the necessity of our filtering phase because if a model cannot correctly assert the behavior of the original algorithm, its performance on a changed version cannot yield reliable insights.
\end{itemize}

% Across the strictly retained programs, LLMs form a strong reference point against which all subsequent semantic-altering and semantic-preserving changes are systematically compared.

\begin{tcolorbox}[colback=gray!10, colframe=gray!50, arc=2mm, boxrule=0.5pt, left=4pt, right=4pt, top=4pt, bottom=4pt]
\textbf{Key Takeaway:} LLMs struggle to generate syntactically valid tests because the task requires deep code comprehension, not just code generation. Consequently, models perform drastically better in Java, where strong typing reduces ambiguity compared to Python. 
\end{tcolorbox}

\subsection{RQ2: Impact of Semantic-Altering Code Changes}
For each program with SAC, we prompt the LLMs in an isolated instance to generate test suites for the updated code and evaluate their effectiveness directly on the changed program.
Across all evaluated programs, test generation effectiveness deteriorates noticeably once program behavior changes. The average test pass rate drops from 100\% in the baseline to 66.5\%, representing a 33.4 percentage-point decrease. Coverage also declines substantially. Line coverage decreases from 79.3\% to 67.4\% (an 11.9 percentage-point reduction, or 15.0\% decrease), while branch coverage falls from 76.1\% to 60.6\% (a 15.5 percentage-point reduction, or 20.4\% decrease). These declines indicate that once program behavior diverges from the original version, LLM-generated tests frequently fail to capture the updated logic and execute fewer relevant program paths.

Beyond the overall drop in performance, several trends emerge. First, branch coverage degrades more than line coverage. This suggests that LLMs struggle particularly with reasoning about altered control-flow decisions introduced by functional changes, especially when the semantic-altering changes occur directly within branch predicates. While the generated tests may still execute parts of the program, they often fail to construct inputs that explore newly introduced branches or modified conditional logic.

Second, we observe a phenomenon we term Scattershot Testing, where the total number of generated tests actually \emph{increases} slightly (from an average of 13.2 to 13.9) despite a severe drop in coverage. This suggests that when LLMs encounter altered logic that diverges from recognized program patterns, they lose their ability to conceptualize an optimal testing strategy. Instead of targeting new execution paths, the models attempt to brute-force coverage by generating a higher volume of shallow tests. This scattershot approach artificially inflates the test count but fails to engage with the altered control flow, perfectly explaining the simultaneous rise in test volume and sharp decline in branch coverage.

\begin{figure*}[t]
    \vspace{-2ex}
    \centering
    \includegraphics[width=1\textwidth,keepaspectratio]{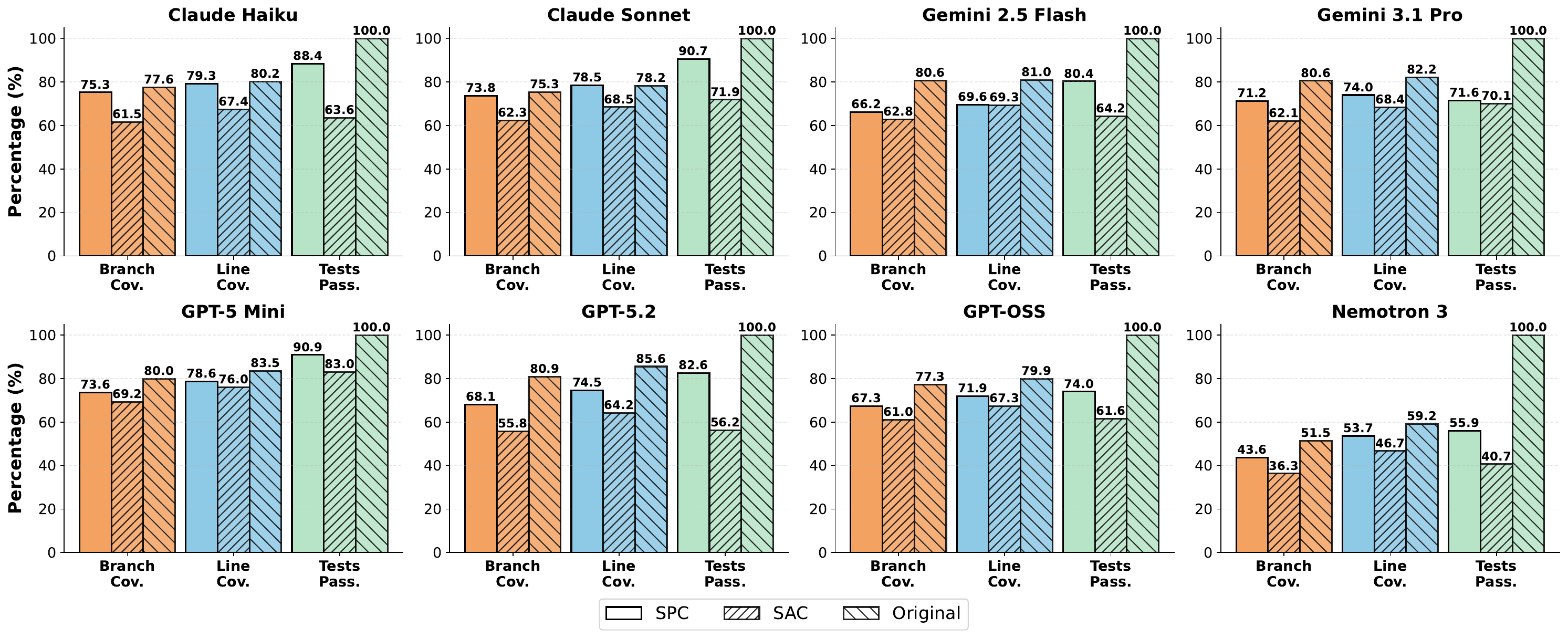}
    \vspace{-4ex}
    \caption{Test pass rates and branch coverage across the original baseline, SPCs, and SACs, broken down by evaluated LLM.}
    \label{fig:aggregate-comparison}
    \vspace{-4ex}
\end{figure*}

Figure~\ref{fig:sam-coverage} further illustrates these patterns across different categories of semantic-altering code changes. Boundary shifts retain notably higher line coverage (73.4\%) compared to other mutations, indicating that the generated tests still manage to execute portions of the affected loops or bounds. However, a persistent gap exists: branch coverage remains substantially lower than line coverage across every single SAC category. This reinforces the observation that while LLMs successfully generate inputs to trigger the altered code blocks along primary execution paths, they struggle to provide the diverse test cases necessary to explore all alternative branches and conditional outcomes of the new logic.

% Taken together, these observations suggest that when program functionality changes, LLM-generated tests do not reliably adapt to the new behavior. Instead of expanding their exploration to cover newly introduced execution paths, the generated suites often remain partially aligned with the original program structure, leading to lower coverage and increased test failures despite generating the same or a high volume of tests.

\begin{tcolorbox}[colback=gray!10, colframe=gray!50, arc=2mm, boxrule=0.5pt, left=4pt, right=4pt, top=4pt, bottom=4pt]
\textbf{Key Takeaway:} Even small functional changes can hurt test performance. Models struggle with new control-flow paths and tend to generate tests based on the old program structure. 
\end{tcolorbox}
\subsection{RQ3: Impact of Semantic-Preserving Code Changes}

As with SACs, we regenerate test suites after applying SPC and evaluate them on the refactored code. Although the underlying semantics remain identical to the baseline, we still observe a measurable decline in test generation performance. Compared to the baseline, the average test pass rate drops from 100\% down to 78.9\%, representing a 21.0 percentage-point decrease. Coverage metrics exhibit a corresponding degradation. Average line coverage decreases from 79.3\% to 73.7\% (a 5.6 percentage-point reduction), while branch coverage falls from 76.1\% to 69.2\% (a 6.9 percentage-point reduction). Although these drops are smaller than those observed under semantic-altering changes, they are alarming given that the execution behavior has not changed at all.

Several patterns emerge from these results. First, branch coverage again declines more than line coverage, suggesting that even purely structural refactorings disrupt the models’ ability to reason about control-flow exploration. While the generated tests still execute many statements in the program, they frequently fail to construct inputs that exercise alternative decision paths.
Second, in our previous analysis of semantic-altering changes, models overcompensated by generating \emph{more} tests. Conversely, under semantic-preserving changes, the average number of generated tests \emph{drops} from 13.2 down to 12.1. This truncation is most extreme under the \emph{Unused Parameter} change, where the models generate an average of only 8.1 tests, causing line coverage to plummet to 65.7\%. This suggests a "distraction effect", when presented with noisy or extraneous structural elements, the model expends its reasoning capacity trying to interpret the dead logic, resulting in an abruptly truncated and incomplete test suite.

These observations show that LLMs' test generation is largely influenced by the syntactic “surface area” of a modification rather than its semantic effect. Larger syntactic changes with no semantic impact (e.g., SPCs) cause substantial deviations in generated tests, while low-syntax, high-semantic changes (e.g., SACs) often go unnoticed. As illustrated in Figure~\ref{fig:spm-coverage}, SPCs consistently reduce branch coverage more than line coverage, highlighting the model’s sensitivity to structural noise rather than true program semantics.

\begin{tcolorbox}[colback=gray!10, colframe=gray!50, arc=2mm, boxrule=0.5pt, left=4pt, right=4pt, top=4pt, bottom=4pt]
\textbf{Key Takeaway:} LLM-based test generation is sensitive to the syntactic and lexical footprint of code changes rather than their true semantic impact.
\end{tcolorbox}

\subsubsection{Discussion: Code Change Sensitivity Across LLMs}

Figure~\ref{fig:aggregate-comparison} provides a consolidated, model-specific view of how different types of code changes impact LLM-generated tests. 
%By evaluating the performance of individual models across the baseline (100\% pass rate), Semantic-Preserving Changes, and Semantic-Altering Changes, two striking themes emerge regarding model resilience and brittleness.
First, the aggregate data confirms a universal hierarchy of degradation where behavioral changes or SACs predictably cause the most catastrophic failures across all models, and purely structural refactorings or SPCs consistently induce a degradation, given that the code is functionally the same. No model evaluated was completely immune to the structural noise introduced by SPCs. Even state-of-the-art models like Claude 4.6 Sonnet and GPT-5 Mini experienced roughly a 10 percentage-point drop in pass rates due solely to structural refactoring. This reinforces the takeaway that current LLM test generation relies heavily on surface-level pattern matching rather than robust semantic understanding.

Second, the breakdown reveals profound differences in resilience. We observe three distinct profiles of code-change sensitivity:

\noindent \textbf{Highly Resilient (e.g., GPT-5 Mini, Claude 4.6 Sonnet):} These models demonstrate the strongest adaptability. GPT-5 Mini, in particular, maintained an 82.9\% pass rate under SACs and a 90.9\% pass rate under SPCs, the highest retention among all evaluated models. They exhibit a stronger capacity to update their internal context when prompted with altered code.

\noindent \textbf{High Baseline but Higher Sensitivity (e.g., GPT-5.2):} Surprisingly, while GPT-5.2 achieved the highest baseline line coverage (85.6\%), it proved fragile to semantic evolution. Its pass rate dropped by nearly 44 percentage points (down to 56.2\%) when faced with SACs. This suggests a severe degree of "memorization over-fitting." 

\noindent\textbf{Size Bottleneck (e.g., Nemotron-3-Nano):} Smaller models struggle with evolutionary context. Nemotron-3-Nano's pass rate dropped to 40.7\% under SACs and 55.9\% under SPCs, showing that limited-parameter models lack the reasoning depth to handle even minor code changes reliably.

Comparing these conditions side-by-side reveals a critical limitation in AI-assisted software engineering. While LLMs are highly proficient at two-shot test generation for public benchmarks, their sensitivity to both behavioral shifts and syntactic noise makes them highly unreliable companions for continuous integration environments where code is constantly evolving and refactoring.

\subsection{RQ4: Failure Attribution under SACs}
\begin{figure}[t]
    % \vspace{-2ex}
    \centering
    \includegraphics[width=0.9\columnwidth,keepaspectratio]{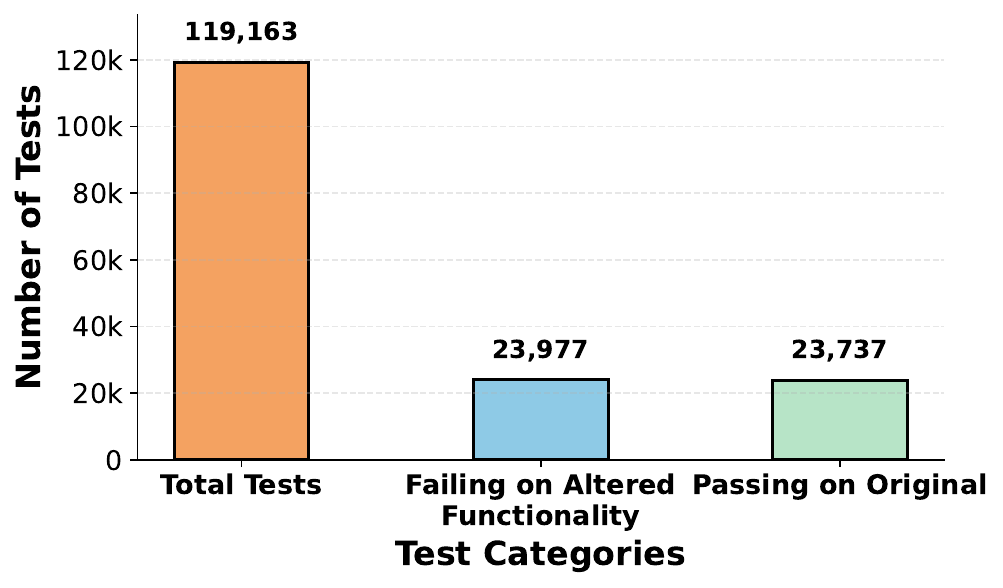}
    \vspace{-3ex}
    \caption{Failure analysis of generated tests under SACs}
    \label{fig:rq6_failure_attribution}
    \vspace{-2ex}
\end{figure}

Figure~\ref{fig:rq6_failure_attribution} disambiguates the cause of test failures observed under semantic-altering code changes. For each change category, the figure reports the total number of generated tests, the number of tests failing on the modified program, and the subset of failing tests that pass when re-executed on the original program.

Across all categories, we analyze 119,163 generated tests, of which 23,977 fail when executed on the modified programs. When these failing tests are re-executed on the corresponding original programs, 23,737 tests pass while still executing the modified code region, yielding a residual-alignment rate above 99\%.

This pattern is consistent across all semantic-altering change types, including arithmetic changes, logical condition changes, boundary shifts, argument swaps, and variable role rebinding, with attribution rates ranging from 98.6\% to 100\%. The near-perfect recovery of failing tests on the original code indicates that the tests themselves are not malformed; instead, they remain aligned with the behavior of the original implementation.

% These results clarify the degradation observed in earlier sections. When program behavior changes, LLM-generated tests rarely adjust their assertions to reflect the new logic. Instead, they continue encoding expectations consistent with the original implementation, which leads to failures when executed against the updated program.

\begin{tcolorbox}[colback=gray!10, colframe=gray!50, arc=2mm, boxrule=0.5pt, left=4pt, right=4pt, top=4pt, bottom=4pt]
\textbf{Key Takeaway:} LLM-generated tests remain aligned to the original code rather than adapting to updated logic.
\end{tcolorbox}

\subsection{RQ5: Regression Awareness}
\label{rq5results}

\begin{figure}[t]
    \vspace{-2ex}
    \centering
    \includegraphics[width=0.9\columnwidth,keepaspectratio]{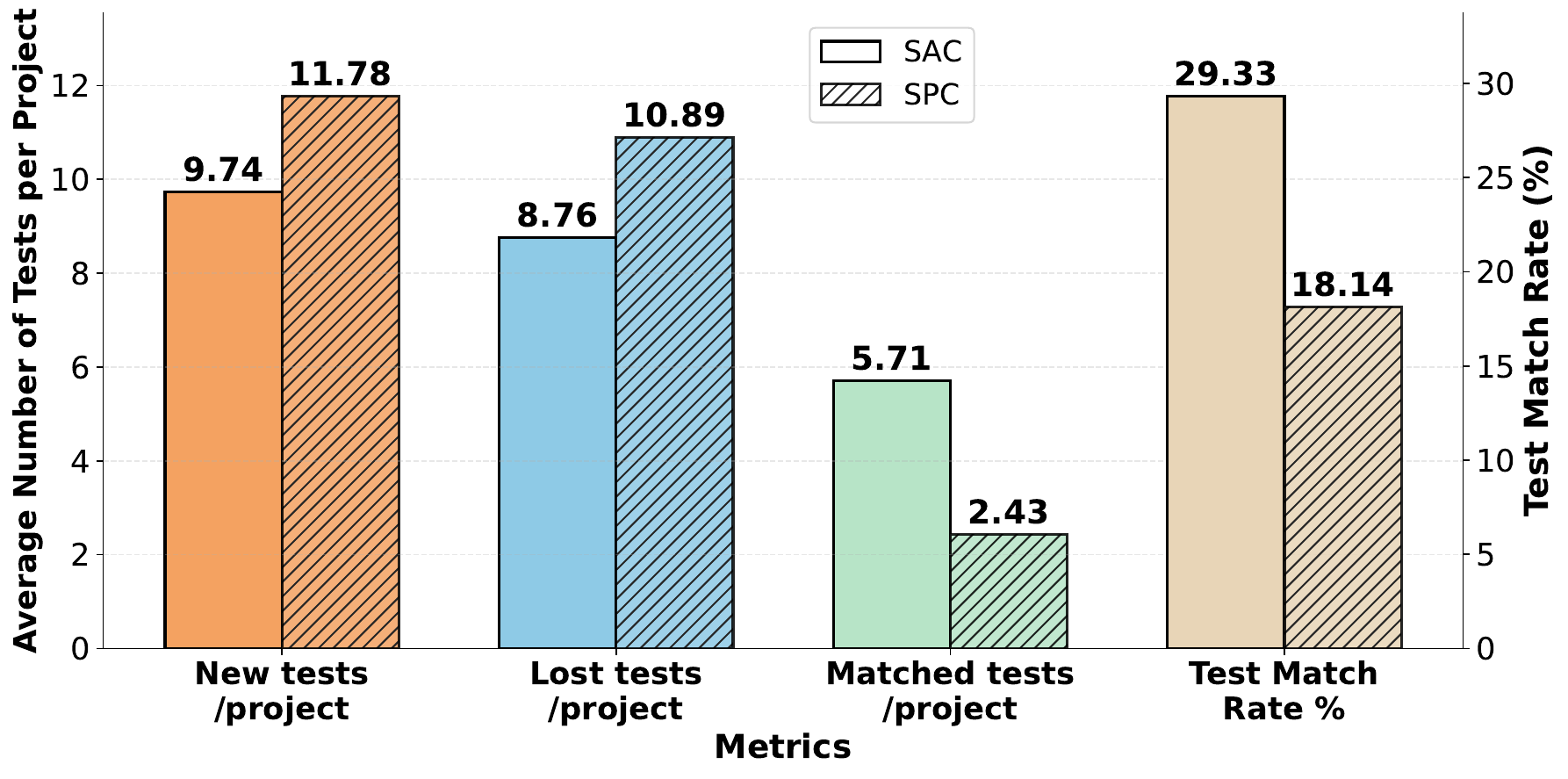}
    \vspace{-3ex}
    \caption{Regression awareness of LLM-generated test suites.}
    \label{fig:rq4_regression_awareness}
    \vspace{-2ex}
\end{figure}

%Beyond coverage degradation, we investigate whether LLM-generated tests exhibit \emph{regression awareness} during code evolution. 
Ideally, when code evolves, a test generator should preserve previously valid, high-quality testing behavior and only add or modify tests necessary to exercise the updated behavior. To analyze this property, we track test-suite continuity across program versions by identifying matched, newly generated, and lost tests.  We use the coverage profiles of two test cases to determine whether they are matched, i.e., they exercise the same lines of code. 

\subsubsection{Test Suite Evolution under Code Changes}

Figure~\ref{fig:rq4_regression_awareness} summarizes how test suites evolve under both SACs and SPCs. We also measured the number of lines modified by each change category. Semantic-altering changes modify an average of 1.0 line per program, while semantic-preserving changes modify 2.4 lines on average, with some changes affecting up to 7 lines. SPCs typically introduce larger structural edits that may cause the LLM to treat the altered program as a different artifact, increasing the likelihood that fewer tests are reused for semantic-preserving changes.

Across 8,585 SAC-enabled programs, we observe high levels of test suite instability. When functionality changes, LLMs generate an average of 9.7 entirely new tests while simultaneously discarding 8.8 previously valid tests, resulting in a test-suite churn of 18.5 tests per evaluation. Consequently, the average test match rate between the baseline and the mutated code is a mere 29.3\%. This indicates that when program logic shifts, models fail to adapt existing tests, opting instead to regenerate the majority of the suite from scratch. As a result, many high-coverage tests are lost, and the newly generated tests achieve lower overall coverage, indicating that LLMs produce less effective tests than prior high-value tests.

Alarmingly, this instability worsens under 10,563 SPC evaluations. Although semantic-preserving changes leave the underlying program semantics unchanged, LLMs exhibit even higher churn (22.7 tests per evaluation), generating 11.8 new tests and discarding 10.9 existing ones.  This drives the average match rate down to a  18.1\%. And since the overall coverage of the regenerated test suite after SPCs always decreases, the discarded tests were more coverage-effective than the newly generated ones.

\begin{tcolorbox}[colback=gray!10, colframe=gray!50, arc=2mm, boxrule=0.5pt, left=4pt, right=4pt, top=4pt, bottom=4pt]
\textbf{Key Takeaway:} LLMs rely on syntactic cues during code changes; they often regenerate test suites from scratch, discarding high-coverage tests and replacing them with lower-coverage ones.

%of incrementally adapting existing tests, models frequently discard valid tests and regenerate new suites, even when program behavior remains unchanged.
\end{tcolorbox}

% \begin{tcolorbox}[colback=gray!10, colframe=gray!50, arc=2mm, boxrule=0.5pt, left=4pt, right=4pt, top=4pt, bottom=4pt]
% \textbf{Key Takeaway:} LLM-generated test suites exhibit limited regression awareness. When code changes occur, LLMs primarily rely on syntactic cues, often generating test suites from scratch, discarding high-coverage tests, and replacing them with lower-coverage ones.
% \end{tcolorbox}
\section{Discussion}
% \gulzar{This section looks more like a summary of key takeaways, We need to say something else here. First, mention two key insights we find (1) LLMs rely on syntactic cues and not semantc impact and (2) the large surface-level changes enouragea LLM to try to understand code logic and not pattern match with code from training time. Then write that we porpose, advancements by preprocessing code before sending to LLMs. One option is to do a code diff and accentuate that code diff to let LLMs know that the change is here and understand the semantic impact. Second, prior work on code change summaries in text, can help LLM understand the nature of code change. Third , static program analysis to extract control and data flow changes  can also help be given to LLM to high the nature and impact of changes. Essntially we need a method that help LLM understand the magnitude of change.  }\sabaat{Addressed}
The results of our study reveal two key insights about the behavior of LLM-based test generation under evolving code. First, the models rely heavily on surface-level syntactic cues rather than the true semantic impact of a change. As a result, even when program behavior remains unchanged, structural refactorings can significantly disrupt the generated test suites. Second, larger surface-level edits appear to push the models away from simple pattern matching and encourage them to attempt deeper reasoning about the code logic. This suggests that the magnitude of code changes strongly influences how LLMs interpret and respond to modified programs.

These observations point toward potential directions for improving the LLM-driven test generation. One promising approach is to preprocess code changes before presenting them to the model. For example, computing a code diff and explicitly highlighting the modified regions could help guide the model's attention toward the parts of the program. Prior work on generating natural language summaries of code diffs and automated commit messages~\cite{lin2024automated, zhang2024using} may also help communicate the intent behind modifications~\cite{10.1145/3728949}, enabling the model to better understand how program behavior has evolved. Additionally, static program analysis techniques could extract structural information such as control-flow and data-flow differences between program versions, providing a clearer signal about the functional impact of the change.
Overall, these directions suggest that future systems should incorporate mechanisms that help LLMs reason about the magnitude and nature of code changes rather than relying solely on raw source code inputs.
\section{Threats to Validity}

\textbf{Internal Validity:} To mitigate the risk of adding unintended functional shifts during SPCs, we test updated code against the original baseline test suites and verify a 100\% pass rate on the original code, ensuring that SPCs are behavior-neutral and do not cause any functional shifts. Another threat is the inherent non-determinism of Large Language Models. Model outputs can vary across attempts, potentially affecting test pass rates and coverage metrics. We addressed this by using fresh, independent model instances for each generation task to prevent context leakage, strictly evaluating immediate outputs to capture the models' baseline reasoning.

\textbf{External Validity:}  Our evaluation is grounded in 22,374 program variants sourced from the Project CodeNet dataset, focusing specifically on Java and Python implementations. While these represent widely used languages and common algorithmic tasks, the findings may not fully generalize to highly complex, multi-file enterprise codebases where deep contextual dependencies exist. However, because the models already struggle to maintain semantic grounding on these relatively simple, self-contained programs, we consider our observed performance degradation to be a conservative baseline, introducing greater architectural complexity would likely only further reduce test reliability. Additionally, our study evaluates a specific subset of 8 state-of-the-art LLMs. As model architectures rapidly evolve, future iterations could display different sensitivities to code evolution.

\textbf{Construct Validity:} We relied on line coverage, branch coverage, and test pass rates to evaluate test suite quality. While these are standard test quality metrics, they do not fully capture a test suite's fault-finding capability or the exact developer intent. However, in the context of evaluating semantic grounding and residual alignment, these metrics, combined with our failure attribution analysis, provide a robust and objective framework for quantifying the models' sensitivity to functional and structural changes.
\section*{Related Work}
\noindent\textbf{Traditional automated test generation.}
Automated test generation is extensively studied with traditional techniques like feedback-directed random testing, such as Randoop~\cite{pacheco2007randoop} and JCrasher~\cite{csallner2004jcrasher}, search-based software testing (SBST), such as EvoSuite~\cite{EvoSuite} and Pynguin~\cite{lukasczyk2020pynguin}, and symbolic execution frameworks like KLEE~\cite{cadar2008klee} and DART~\cite{godefroid2005dart}. These approaches generate test inputs to maximize code coverage, such as branches or statements. 
%For instance, EvoSuite employs genetic algorithms to evolve test suites that traverse complex execution paths, while Randoop incrementally builds sequences of method calls to uncover contract violations. 
Most existing approaches are coupled with specific programming languages and require substantial engineering effort to design and maintain test generation frameworks. Despite this complexity, they lack many features necessary for automated and efficient test generation.
%Despite their success in achieving high structural coverage, these traditional tools face fundamental limitations regarding semantic intent. Empirical evaluations, such as the large-scale studies by Shamshiri et al.~\cite{shamshiri2015evosuite} and Almasi et al.~\cite{almasi2017industrial}, reveal that while algorithmically generated tests excel at finding crash-inducing inputs, they struggle to produce meaningful, human-readable assertions or capture the true functional logic of the software. Because they rely entirely on the static structure of the codebase, they cannot infer underlying developer intent. To bridge this semantic gap and generate developer-aligned, logically grounded test suites, the software engineering community has increasingly turned toward learning-based paradigms, culminating in the rapid adoption of LLMs for automated test generation.

\noindent\textbf{LLM-based automated test generation.}
Recent work has explored using LLMs for test generation with some proposing prompting strategies and fine-tuning to generate developer-like unit tests that achieve both high structural coverage and semantic readability~\cite{lahiri2022interactive, lemieux2023codamosa}. For example, Sch{"a}fer et al.~\cite{schafer2023testpilot} and Lops et al.~\cite{10962454} provide comprehensive empirical evaluations and generation systems for LLM-driven unit testing. Further advancing these generation techniques, SymPrompt~\cite{symprompt2024} uses a multi-stage, code-aware prompting process aligned with execution paths, while IntUT~\cite{intut2025} uses explicit test intentions (e.g., inputs, mocks, and expected outcomes) to guide generation. Similarly, Yuan et al.~\cite{yuan2023chatunitest} proposed ChatUniTest, an automated framework leveraging adaptive focal context to minimize hallucinations when generating test suites for complex Java projects. 
%Recent applications also demonstrate the scalability of LLMs in dynamic environments; 
Meta's unit test improvement framework~\cite{testgen2024} improves and extends existing tests in industrial workflows, and E-Test~\cite{etest2026} continuously augments test suites using execution scenarios harvested from production logs. Extensive benchmarking efforts have also sought to quantify these generation capabilities. Siddiq et al.~\cite{siddiq2023exploring}, Yuan et al.~\cite{yuan2023evaluating}, and Jiang et al.~\cite{10.1145/3691620.3695513} evaluated various LLMs on their baseline ability to generate unit tests. Despite these promising results, current evaluations predominantly rely on well-known, publicly available benchmarks. Existing studies fail to isolate this potential memorization from true semantic reasoning. 

%Because these static datasets are highly likely to be included in the pre-training corpora of large language models, achieving high test generation coverage on them does not guarantee genuine code comprehension. The models might simply be reproducing test suites they have already been exposed to during training. Our work directly addresses this gap. By systematically altering the underlying code of these benchmark programs, we force the models to generate tests for newly modified logic. This approach allows us to rigorously measure whether they truly understand the provided code or are merely regurgitating exposed training patterns.

\noindent\textbf{Code modifications and LLM robustness.}
%A distinct line of research examines how neural models respond to structural and lexical code modifications. Researchers have increasingly scrutinized the 
The fragility of LLMs' generative capability is also studied when presented with variations of standard programming tasks. For instance, Wang et al.~\cite{wang2023recode} introduced ReCode, a framework for evaluating the robustness of code generation models against semantic-preserving mutations. Similarly, Rabin et al.~\cite{rabin2021generalizability}, and Yefet et al.~\cite{yefet2020adversarial} investigated the generalizability of neural program analyzers under semantic-preserving mutations, demonstrating that simple changes like variable renaming can severely degrade model performance. Other empirical studies, such as those by Dong et al.~\cite{dong2023understanding} and Yang et al.~\cite{yang2022natural}, assess how adversarial refactoring and natural mutations impact downstream tasks like vulnerability detection and code summarization. These works primarily evaluate whether a model can still produce the correct code snippet or identify a bug when the code or prompt is altered, mostly testing LLM generative capabilities. They do not evaluate the LLM code comprehension ability needed for test generation. 

%explore how functional updates to publicly available code or non-functional refactoring affect the model's performance during test generation.

\noindent\textbf{Memorization and semantic grounding.}
A fundamental challenge in evaluating Large Language Models is distinguishing genuine reasoning from the memorization of pre-training data. Carlini et al.~\cite{carlini2021extracting} and Lee et al.~\cite{lee2022deduplicating} demonstrate that LLMs can exactly reproduce substantial portions of their training corpora. Recent surveys and contamination-focused studies~\cite{yang2024data,dong2024generalization} also highlight how benchmark contamination can inflate the perceived performance of code generation models. %To mitigate this reliance on memorized patterns, dynamic evaluation frameworks like DyCodeEval~\cite{chen2025dycodeeval} construct fresh programming tasks to ensure models are tested on unseen logic.
% \vspace{-0.5em}
\section{Conclusion}
We examined how code changes affect the reliability of LLM-based test generation. Using an automated mutation-driven evaluation framework on over 22,374 program variants, we evaluate model behavior under both semantic-altering and semantic-preserving changes. While LLMs achieve strong baseline performance on unmodified programs (79.3\% line and 76.1\% branch coverage with fully passing tests), test quality degrades substantially once code changes are introduced. Under semantic-altering changes, many generated tests remain aligned with the original program behavior, with over 99\% of failing tests passing on the original code. Under semantic-preserving changes, coverage and pass rates decline despite unchanged functionality. These results suggest that current LLM-based test generation fails to reason about the semantic impact of code changes and instead responds mainly to the magnitude of syntactic differences in the code.

\balance
\bibliographystyle{ACM-Reference-Format}
\bibliography{main}

@inproceedings{Fuzz4All,
author = {Xia, Chunqiu Steven and Paltenghi, Matteo and Le Tian, Jia and Pradel, Michael and Zhang, Lingming},
title = {Fuzz4All: Universal Fuzzing with Large Language Models},
year = {2024},
isbn = {9798400702174},
publisher = {Association for Computing Machinery},
address = {New York, NY, USA},
url = {https://doi.org/10.1145/3597503.3639121},
doi = {10.1145/3597503.3639121},
booktitle = {Proceedings of the IEEE/ACM 46th International Conference on Software Engineering},
articleno = {126},
numpages = {13},
location = {Lisbon, Portugal},
series = {ICSE '24}
}

@article{Fan2023LargeLM,
  title={Large Language Models for Software Engineering: Survey and Open Problems},
  author={Angela Fan and Beliz Gokkaya and Mark Harman and Mitya Lyubarskiy and Shubho Sengupta and Shin Yoo and Jie M. Zhang},
  journal={2023 IEEE/ACM International Conference on Software Engineering: Future of Software Engineering (ICSE-FoSE)},
  year={2023},
  pages={31-53},
  url={https://api.semanticscholar.org/CorpusID:263671720}
}

@ARTICLE{2026arXiv260212256C,
       author = {{Chudic}, Alex and {{\c{C}}al{\i}kl{\i}}, G{\"u}l},
        title = "{Automated Test Suite Enhancement Using Large Language Models with Few-shot Prompting}",
      journal = {arXiv e-prints},
         year = 2026,
        month = feb,
          eid = {arXiv:2602.12256},
        pages = {arXiv:2602.12256},
          doi = {10.48550/arXiv.2602.12256},
archivePrefix = {arXiv},
       eprint = {2602.12256},
 primaryClass = {cs.SE},
       adsurl = {https://ui.adsabs.harvard.edu/abs/2026arXiv260212256C}
}

@inproceedings{EvoSuite,
author = {Fraser, Gordon and Arcuri, Andrea},
title = {EvoSuite: automatic test suite generation for object-oriented software},
year = {2011},
isbn = {9781450304436},
publisher = {Association for Computing Machinery},
address = {New York, NY, USA},
url = {https://doi.org/10.1145/2025113.2025179},
doi = {10.1145/2025113.2025179},
booktitle = {Proceedings of the 19th ACM SIGSOFT Symposium and the 13th European Conference on Foundations of Software Engineering},
pages = {416–419},
numpages = {4},
location = {Szeged, Hungary},
series = {ESEC/FSE '11}
}

@inproceedings{10.1145/3597503.3639219,
author = {Du, Xueying and Liu, Mingwei and Wang, Kaixin and Wang, Hanlin and Liu, Junwei and Chen, Yixuan and Feng, Jiayi and Sha, Chaofeng and Peng, Xin and Lou, Yiling},
title = {Evaluating Large Language Models in Class-Level Code Generation},
year = {2024},
isbn = {9798400702174},
publisher = {Association for Computing Machinery},
address = {New York, NY, USA},
url = {https://doi.org/10.1145/3597503.3639219},
doi = {10.1145/3597503.3639219},
booktitle = {Proceedings of the IEEE/ACM 46th International Conference on Software Engineering},
articleno = {81},
numpages = {13},
location = {Lisbon, Portugal},
series = {ICSE '24}
}

@article{puri2021codenet,
  title={Project CodeNet: A Large-Scale AI for Code Dataset for Learning a Diversity of Coding Tasks},
  author={Puri, Ruchir and Kung, David S and Janssen, Gaetan and Zhang, Wei and Bury, Giacomo and Griguer, Jason and Fanebust, Luca and Dykeman, Bryce and Shah, Tanveer and Kamani, Rishit and others},
  journal={arXiv preprint arXiv:2105.12655},
  year={2021}
}

@inproceedings{pacheco2007randoop,
  title={Feedback-directed random test generation},
  author={Pacheco, Carlos and Lahiri, Shuvendu K and Ernst, Michael D and Ball, Thomas},
  booktitle={Proceedings of the 29th international conference on Software Engineering (ICSE)},
  pages={75--84},
  year={2007},
  organization={IEEE}
}

@inproceedings{cadar2008klee,
  title={KLEE: Unassisted and Automatic Generation of High-Coverage Tests for Complex Systems Programs.},
  author={Cadar, Cristian and Dunbar, Daniel and Engler, Dawson R and others},
  booktitle={USENIX Symposium on Operating Systems Design and Implementation (OSDI)},
  volume={8},
  pages={209--224},
  year={2008}
}

@misc{schafer2023testpilot,
      title={An Empirical Evaluation of Using Large Language Models for Automated Unit Test Generation}, 
      author={Max Schäfer and Sarah Nadi and Aryaz Eghbali and Frank Tip},
      year={2023},
      eprint={2302.06527},
      archivePrefix={arXiv},
      primaryClass={cs.SE},
      url={https://arxiv.org/abs/2302.06527}, 
}

@inproceedings{yuan2023chatunitest,
  title={Chatunitest: a chatgpt-based automated unit test generation framework},
  author={Yuan, Zhiqiang and Bai, Yiling and Huo, Xuan and Chen, Cuiyun and others},
  booktitle={Proceedings of the 32nd ACM SIGSOFT International Symposium on Software Testing and Analysis},
  pages={1--13},
  year={2023}
}

@inproceedings{siddiq2023exploring,
  title={Exploring the effectiveness of large language models in automated unit test generation},
  author={Siddiq, Mohammed Latif and Santos, Joanna CS},
  booktitle={2023 IEEE International Conference on Software Maintenance and Evolution (ICSME)},
  pages={332--342},
  year={2023},
  organization={IEEE}
}

@inproceedings{wang2023recode,
  title={ReCode: Robustness Evaluation of Code Generation Models},
  author={Wang, Shuo and Zheng, Pei and others},
  booktitle={Proceedings of the 61st Annual Meeting of the Association for Computational Linguistics (Volume 1: Long Papers)},
  pages={2634--2649},
  year={2023}
}

@article{rabin2021generalizability,
  title={On the generalizability of neural program models with respect to semantic-preserving transformations},
  author={Rabin, Md Rafiqul Islam and Al-Haj, Amin and Alipour, Amin},
  journal={Information and Software Technology},
  volume={135},
  pages={106552},
  year={2021},
  publisher={Elsevier}
}

@article{dong2023understanding,
  title={Understanding the Robustness of Large Language Models for Code},
  author={Dong, Yihong and Jiang, Junda and others},
  journal={arXiv preprint arXiv:2305.14886},
  year={2023}
}

@inproceedings{yang2022natural,
  title={Natural attack for pre-trained models of code},
  author={Yang, Zhou and Shi, Jieke and He, Junda and Lo, David},
  booktitle={Proceedings of the 44th International Conference on Software Engineering},
  pages={1482--1493},
  year={2022}
}

@inproceedings{carlini2021extracting,
  title={Extracting training data from large language models},
  author={Carlini, Nicholas and Tramer, Florian and Wallace, Eric and Jagielski, Matthew and Herbert-Voss, Ariel and Lee, Katherine and Roberts, Adam and Brown, Tom and Song, Dawn and Erlingsson, Ulfar and others},
  booktitle={30th USENIX Security Symposium (USENIX Security 21)},
  pages={2633--2650},
  year={2021}
}

@misc{dong2024generalization,
      title={Memorize or Generalize? Evaluating LLM Code Generation with Code Rewriting}, 
      author={Lizhe Zhang and Wentao Chen and Li Zhong and Letian Peng and Zilong Wang and Jingbo Shang},
      year={2025},
      eprint={2503.02296},
      archivePrefix={arXiv},
      primaryClass={cs.AI},
      url={https://arxiv.org/abs/2503.02296}, 
}

@article{symprompt2024,
author = {Ryan, Gabriel and Jain, Siddhartha and Shang, Mingyue and Wang, Shiqi and Ma, Xiaofei and Ramanathan, Murali Krishna and Ray, Baishakhi},
title = {Code-Aware Prompting: A Study of Coverage-Guided Test Generation in Regression Setting using LLM},
year = {2024},
issue_date = {July 2024},
publisher = {Association for Computing Machinery},
address = {New York, NY, USA},
volume = {1},
number = {FSE},
url = {https://doi.org/10.1145/3643769},
doi = {10.1145/3643769},
journal = {Proc. ACM Softw. Eng.},
month = jul,
articleno = {43},
numpages = {21}
}

@inbook{intut2025,
author = {Nan, Zifan and Guo, Zhaoqiang and Liu, Kui and Xia, Xin},
title = {Test Intention Guided LLM-Based Unit Test Generation},
year = {2025},
isbn = {9798331505691},
publisher = {IEEE Press},
url = {https://doi.org/10.1109/ICSE55347.2025.00243},
booktitle = {Proceedings of the IEEE/ACM 47th International Conference on Software Engineering},
pages = {1026–1038},
numpages = {13}
}

@inproceedings{testgen2024,
author = {Alshahwan, Nadia and Chheda, Jubin and Finogenova, Anastasia and Gokkaya, Beliz and Harman, Mark and Harper, Inna and Marginean, Alexandru and Sengupta, Shubho and Wang, Eddy},
title = {Automated Unit Test Improvement using Large Language Models at Meta},
year = {2024},
isbn = {9798400706585},
publisher = {Association for Computing Machinery},
address = {New York, NY, USA},
url = {https://doi.org/10.1145/3663529.3663839},
doi = {10.1145/3663529.3663839},
booktitle = {Companion Proceedings of the 32nd ACM International Conference on the Foundations of Software Engineering},
pages = {185–196},
numpages = {12},
location = {Porto de Galinhas, Brazil},
series = {FSE 2024}
}

@misc{etest2026,
      title={E-Test: E'er-Improving Test Suites}, 
      author={Ketai Qiu and Luca Di Grazia and Leonardo Mariani and Mauro Pezzè},
      year={2025},
      eprint={2510.19860},
      archivePrefix={arXiv},
      primaryClass={cs.SE},
      url={https://arxiv.org/abs/2510.19860}, 
}

@inproceedings{csallner2004jcrasher,
  title={JCrasher: an automatic robustness tester for Java},
  author={Csallner, Christoph and Smaragdakis, Yannis},
  booktitle={Software: Practice and Experience},
  volume={34},
  number={11},
  pages={1025--1050},
  year={2004},
  publisher={Wiley Online Library}
}

@inproceedings{lukasczyk2020pynguin,
  title={Pynguin: Automated unit test generation for Python},
  author={Lukasczyk, Stephan and Kroi{\ss}, Christian and Fraser, Gordon},
  booktitle={Proceedings of the ACM/IEEE 42nd International Conference on Software Engineering},
  pages={168--179},
  year={2020}
}

@inproceedings{godefroid2005dart,
  title={DART: directed automated random testing},
  author={Godefroid, Patrice and Klarlund, Nils and Sen, Koushik},
  booktitle={Proceedings of the 2005 ACM SIGPLAN conference on Programming language design and implementation},
  pages={213--223},
  year={2005}
}

@article{lahiri2022interactive,
  title={Interactive code generation via test-driven user-intent formalization},
  author={Lahiri, Shuvendu K and others},
  journal={arXiv preprint arXiv:2208.05950},
  year={2022}
}

@inproceedings{lemieux2023codamosa,
  title={CODAMOSA: Escaping Coverage Plateaus in Test Generation with Pre-trained Large Language Models},
  author={Lemieux, Caroline and others},
  booktitle={Proceedings of the 45th International Conference on Software Engineering},
  year={2023}
}

@misc{yuan2023evaluating,
      title={Evaluating Instruction-Tuned Large Language Models on Code Comprehension and Generation}, 
      author={Zhiqiang Yuan and Junwei Liu and Qiancheng Zi and Mingwei Liu and Xin Peng and Yiling Lou},
      year={2023},
      eprint={2308.01240},
      archivePrefix={arXiv},
      primaryClass={cs.CL},
      url={https://arxiv.org/abs/2308.01240}, 
}

@inproceedings{yefet2020adversarial,
  title={Adversarial examples for models of code},
  author={Yefet, Noam and Alon, Uri and Yahav, Eran},
  booktitle={Proceedings of the ACM on Programming Languages},
  volume={4},
  number={OOPSLA},
  pages={1--30},
  year={2020}
}

@inproceedings{lee2022deduplicating,
  title={Deduplicating Training Data Makes Language Models Better},
  author={Lee, Katherine and Ippolito, Daphne and Nystrom, Andrew and Zhang, Chiyuan and Eck, Douglas and Callison-Burch, Chris and Sirivianos, Nicholas},
  booktitle={Proceedings of the 60th Annual Meeting of the Association for Computational Linguistics},
  pages={8424--8445},
  year={2022}
}

@article{yang2024data,
  title={Data Contamination in Large Language Models: A Survey},
  author={Yang, Shuo and others},
  journal={arXiv preprint arXiv:2402.10825},
  year={2024}
}

@inproceedings{kovalenko2023unittestbot,

  title={UnitTestBot: Automated Unit Test Generation for C Code in Integrated Development Environments},

  author={Kovalenko, Vladimir and others},

  booktitle={IEEE/ACM 45th International Conference on Software Engineering: Software Engineering in Practice (ICSE-SEIP)},

  year={2023},

  organization={IEEE}

}

@misc{rho2024tamingbeastfullyautomated,

      title={Taming the Beast: Fully Automated Unit Testing with Coyote C++}, 

      author={Sanghoon Rho and Philipp Martens and Seungcheol Shin and Yeoneo Kim},

      year={2024},

      eprint={2401.01073},

      archivePrefix={arXiv},

      primaryClass={cs.PL},

      url={https://arxiv.org/abs/2401.01073}, 

}

@article{gptoss,
  title={gpt-oss-120b \& gpt-oss-20b Model Card},
  author={{OpenAI}},
  journal={arXiv preprint arXiv:2508.10925},
  year={2025}
}

@article{nemotron3nano,
  title={Nemotron 3 Nano: Open, Efficient Mixture-of-Experts Hybrid Mamba-Transformer Model for Agentic Reasoning},
  author={{NVIDIA}},
  journal={arXiv preprint arXiv:2512.20848},
  year={2025}
}

@article{gpt5,
  title={OpenAI GPT-5 System Card},
  author={Singh, Aaditya and others},
  journal={arXiv preprint arXiv:2601.03267},
  year={2026}
}

@misc{gpt52,
  title={Advancing science and math with GPT-5.2},
  author={{OpenAI}},
  howpublished={\url{https://openai.com/index/gpt-5-2-for-science-and-math/}},
  year={2025}
}

@misc{claude45haiku,
  title={Claude Haiku 4.5 System Card},
  author={{Anthropic}},
  howpublished={\url{https://www.anthropic.com/claude-haiku-4-5-system-card}},
  year={2025}
}

@misc{claude46sonnet,
  title={Claude 4.6 Sonnet System Card},
  author={{Anthropic}},
  howpublished={\url{https://www.anthropic.com/claude-4-6-sonnet-system-card}},
  year={2026}
}

@article{gemini25flash,
  title={Gemini 2.5: Advancing Multimodal Capabilities for Fast Inference},
  author={{Gemini Team}},
  journal={Google DeepMind Technical Report},
  year={2025}
}

@article{gemini31pro,
  title={Gemini 3.1: Next-Generation Reasoning and Agentic Capabilities},
  author={{Gemini Team}},
  journal={Google DeepMind Technical Report},
  year={2026}
}

@article{HumanEval,
  title={Evaluating Large Language Models Trained on Code},
  author={Mark Chen and et al.},
  journal = {arXiv preprint arXiv:2107.03374},
  year={2021},
  eprint={2107.03374},
  archivePrefix={arXiv},
  primaryClass={cs.LG}
}

@misc{MBPP,
      title={Program Synthesis with Large Language Models}, 
      author={Jacob Austin and Augustus Odena and Maxwell Nye and Maarten Bosma and Henryk Michalewski and David Dohan and Ellen Jiang and Carrie Cai and Michael Terry and Quoc Le and Charles Sutton},
      year={2021},
      eprint={2108.07732},
      archivePrefix={arXiv},
      primaryClass={cs.PL},
      url={https://arxiv.org/abs/2108.07732}, 
}

@INPROCEEDINGS{10962454,
  author={Lops, Andrea and Narducci, Fedelucio and Ragone, Azzurra and Trizio, Michelantonio and Bartolini, Claudio},
  booktitle={2025 IEEE International Conference on Software Testing, Verification and Validation Workshops (ICSTW)}, 
  title={A System for Automated Unit Test Generation using Large Language Models and Assessment of Generated Test Suites}, 
  year={2025},
  volume={},
  number={},
  pages={29-36},
  doi={10.1109/ICSTW64639.2025.10962454}}

@inproceedings{10.1145/3691620.3695513,
author = {Jiang, Zongze and Wen, Ming and Cao, Jialun and Shi, Xuanhua and Jin, Hai},
title = {Towards Understanding the Effectiveness of Large Language Models on Directed Test Input Generation},
year = {2024},
isbn = {9798400712487},
publisher = {Association for Computing Machinery},
address = {New York, NY, USA},
url = {https://doi.org/10.1145/3691620.3695513},
doi = {10.1145/3691620.3695513},
booktitle = {Proceedings of the 39th IEEE/ACM International Conference on Automated Software Engineering},
pages = {1408–1420},
numpages = {13},
location = {Sacramento, CA, USA},
series = {ASE '24}
}

@inproceedings{ neurosymbolic-oracles-ase2025,
  author = { Amrit Siddiq and Michael D. Ernst and Mauro Pezzè },
  title = { Do LLMs generate useful test oracles? An empirical study with an unbiased dataset },
  booktitle = { Proceedings of the 40th IEEE/ACM International Conference on Automated Software Engineering (ASE 2025) },
  year = { 2025 },
  note = { To appear }
}

@misc{manes2019artscienceengineeringfuzzing,
      title={The Art, Science, and Engineering of Fuzzing: A Survey}, 
      author={Valentin J. M. Manes and HyungSeok Han and Choongwoo Han and Sang Kil Cha and Manuel Egele and Edward J. Schwartz and Maverick Woo},
      year={2019},
      eprint={1812.00140},
      archivePrefix={arXiv},
      primaryClass={cs.CR},
      url={https://arxiv.org/abs/1812.00140}, 
}

@inproceedings{Nguyen_2025,
   title={Are the Majority of Public Computational Notebooks Pathologically Non-Executable?},
   url={http://dx.doi.org/10.1109/MSR66628.2025.00070},
   DOI={10.1109/msr66628.2025.00070},
   booktitle={2025 IEEE/ACM 22nd International Conference on Mining Software Repositories (MSR)},
   publisher={IEEE},
   author={Nguyen, Tien and Gill, Waris and Gulzar, Muhammad Ali},
   year={2025},
   month=apr, pages={396–407} }

@inproceedings{khajezade2024investigating,
  title={Investigating the Efficacy of Large Language Models for Code Clone Detection},
  author={Khajezade, Mohammad and others},
  booktitle={Proceedings of the 32nd IEEE/ACM International Conference on Program Comprehension (ICPC)},
  year={2024},
  organization={IEEE}
}

@inproceedings{nicoletti2024crosslingual,
  title={Cross-lingual Code Clone Detection: When LLMs Fail Short Against Embedding-based Classifier},
  author={Nicoletti, Luca and others},
  booktitle={Proceedings of the 39th IEEE/ACM International Conference on Automated Software Engineering (ASE)},
  year={2024},
  organization={IEEE}
}

@article{10.1145/3728949,
author = {Fang, Minying and Yuan, Xing and Li, Yuying and Li, Haojie and Fang, Chunrong and Du, Junwei},
title = {Enhanced Prompting Framework for Code Summarization with Large Language Models},
year = {2025},
issue_date = {July 2025},
publisher = {Association for Computing Machinery},
address = {New York, NY, USA},
volume = {2},
number = {ISSTA},
url = {https://doi.org/10.1145/3728949},
doi = {10.1145/3728949},
journal = {Proc. ACM Softw. Eng.},
month = jun,
articleno = {ISSTA072},
numpages = {24}
}

@article{lin2024automated,
  title={Automated Commit Message Generation with Large Language Models: An Empirical Study and Beyond},
  author={Lin, Yucen and others},
  journal={IEEE Transactions on Software Engineering},
  year={2024},
  publisher={IEEE}
}

@misc{zhang2024using,
      title={Using Large Language Models for Commit Message Generation: A Preliminary Study}, 
      author={Linghao Zhang and Jingshu Zhao and Chong Wang and Peng Liang},
      year={2024},
      eprint={2401.05926},
      archivePrefix={arXiv},
      primaryClass={cs.SE},
      url={https://arxiv.org/abs/2401.05926}, 
}

\end{document}